\documentclass[%
 reprint,
 amsmath,amssymb,
 aps,
 prl
]{revtex4-1}
\usepackage{graphicx}
\usepackage{subfigure}
\usepackage{epstopdf}
\usepackage{amsmath}
\usepackage{amssymb}
\usepackage{amsfonts}
\usepackage{mathrsfs}
\usepackage{theorem}
\usepackage{bm}
\usepackage{url}
\usepackage[T1]{fontenc}
\usepackage{csquotes}
\MakeOuterQuote{"}

\usepackage{algorithm}
\usepackage{algorithmicx}
\usepackage{algpseudocode}

\usepackage{dcolumn}
\usepackage{color}

\def\vec#1{\bm{#1}} 

\newcommand{\tr}{\operatorname{tr}}

\newcommand{\rmi}{\mathrm{i}}

\newcommand{\braket}[2]{\ensuremath{\left\langle #1|#2 \right\rangle}}
\newcommand{\ketbra}[2]{\ensuremath{\left| #1 \rangle \langle #2\right|}}

\newcommand{\be}{\begin{equation}}
\newcommand{\ee}{\end{equation}}
\newcommand{\ba}{\begin{align}}
\newcommand{\ea}{\end{align}}

\def\<{\langle}  
\def\>{\rangle}  
\newcommand{\ket}[1]{| #1\>}
\newcommand{\bra}[1]{\< #1|}

\def\eqref#1{\textup{(\ref{#1})}}  

\newcommand{\cref}[1]{Conjecture~\ref{#1}}
\newcommand{\Cref}[1]{Conjecture~\ref{#1}}

\begin{document}



\title{Ultimate precision of multi-parameter quantum magnetometry under the parallel scheme}

\author{Zhibo Hou}
\thanks{These authors contributed equally to this work.}
\affiliation{Key Laboratory of Quantum Information,University of Science and Technology of China, CAS, Hefei 230026, P. R. China}
\affiliation{CAS Center For Excellence in Quantum Information and Quantum Physics}
\author{Hongzhen Chen}
\thanks{These authors contributed equally to this work.}
\affiliation{Department of Mechanical and Automation Engineering, The Chinese University of Hong Kong, Shatin, Hong Kong}
\author{Liqiang Liu}
\thanks{These authors contributed equally to this work.}
\affiliation{Department of Mechanical and Automation Engineering, The Chinese University of Hong Kong, Shatin, Hong Kong}
\author{Zhao Zhang}
\affiliation{Key Laboratory of Quantum Information,University of Science and Technology of China, CAS, Hefei 230026, P. R. China}
\affiliation{CAS Center For Excellence in Quantum Information and Quantum Physics}
\author{Guo-Yong Xiang}
\email{gyxiang@ustc.edu.cn}
\affiliation{Key Laboratory of Quantum Information,University of Science and Technology of China, CAS, Hefei 230026, P. R. China}
\affiliation{CAS Center For Excellence in Quantum Information and Quantum Physics}
\author{Chuan-Feng Li}
\affiliation{Key Laboratory of Quantum Information,University of Science and Technology of China, CAS, Hefei 230026, P. R. China}
\affiliation{CAS Center For Excellence in Quantum Information and Quantum Physics}
\author{Guang-Can Guo}
\affiliation{Key Laboratory of Quantum Information,University of Science and Technology of China, CAS, Hefei 230026, P. R. China}
\affiliation{CAS Center For Excellence in Quantum Information and Quantum Physics}
\author{Haidong Yuan}
\email{hdyuan@mae.cuhk.edu.hk}
\affiliation{Department of Mechanical and Automation Engineering, The Chinese University of Hong Kong, Shatin, Hong Kong}

\date{\today}

\begin{abstract}
The precise measurement of a magnetic field is one of the most fundamental and important tasks in quantum metrology. Although extensive studies on quantum magnetometry have been carried out over past decades, the ultimate precision that can be achieved for the estimation of all three components of a magnetic field with entangled probe states under the parallel scheme remains unknown. Here we present the ultimate lower bound for the sum of arbitrarily weighted variances in the estimation of all three components of a magnetic field under the parallel scheme and show that this lower bound can be achieved for sufficiently large $N$. The optimal entangled probe state that achieves the ultimate precision is also explicitly constructed. The obtained precision sets the ultimate limit for the multi-parameter quantum magnetometry under the parallel scheme, which is of fundamental interest and importance in quantum metrology. Our approach also provides a way to characterize the tradeoff among the precisions of multiple parameters that arise from the constraints on the probe states.
\end{abstract}





\maketitle

\begin{figure}[t]
  \centering
  \includegraphics[width=0.48\textwidth]{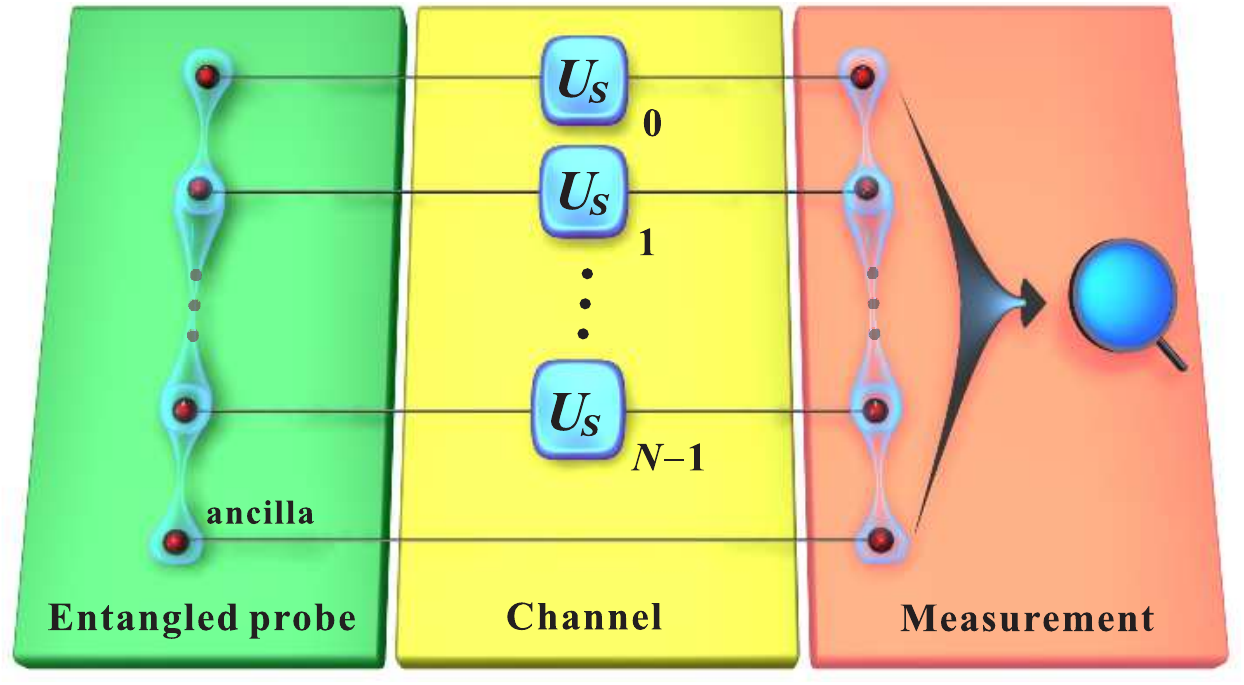}
	\caption{\label{fig:scheme} Parallel scheme for multi-parameter quantum magnetometry. Here $U_s=e^{-i\vec{B}\cdot\vec{\sigma}t}$ describes the unitary dynamics on each of the $N$ spin due to the interaction between the spin and the magnetic field. An additional ancillary system can be used.    }
\end{figure}

Many applications of quantum metrology can be reduced to the measurement and estimation of a magnetic field. For example, various applications in quantum bio-sensing with NV-centers are achieved by measuring the magnetic field of the targeted bio-molecules\cite{Christian2014}. Quantum magnetometry under the parallel scheme that utilizes entangled probe states, as shown in Fig.\ref{fig:scheme}, has been studied over many decades since the pioneer work of Helstrom and Holevo\cite{HELS67,HOLE82}. The ultimate precision, however, is only well understood for the single-parameter quantum magnetometry. An example extensively studied is the estimation of the Z-component of a magnetic field, i.e., the estimation of the projection of the magnetic field on the Z-axis. In this case, the ultimate precision for the local estimation, where the experiment needs to be repeated for sufficient number of times, is achieved by the GHZ-type state as $\frac{|00\cdots 0\rangle+|11\cdots 1\rangle}{\sqrt{2}}$, under which the variance of the estimation scales as $\frac{1}{N^2}$\cite{GIOV04,GIOV06}. For the Bayesian estimation, where the experiment is only performed once, the minimal Holevo covariance is achieved with the Berry-Wisemen type of states\cite{Berry2000}. For the estimation of all three components of the magnetic field, the answer is only known for special cases. For the Bayesian estimation, the optimal performance for the estimation of the generated unitary rotation has been studied under the assumption of uniform prior distribution\cite{Acin01PRA,Mauro2001,Chiribella2004,Chiribella2005,Peres2001,Bagan2004,Bagan20042}. For the local estimation, the optimal performance is only known when the unitary rotation generated is close to the Identity operator and the figure of merit is taken as the sum of equally weighted variance under some specific parametrization\cite{Piotr2008,Datta2016,Manuel2004,Fujiwara2001,Imai2007,Manuel20042,Hayashi2006,HAYASHI2006183,Manuel2005}. For general unitary rotations, a heuristic state is provided in \cite{Datta2016} with the achieved precision matching the optimal performance in the weak limit, i.e., when the magnetic field is close to $0$ and the generated unitary is close to the Identity operator. In general, however, the optimal performance of the multi-parameter quantum magnetometry under the parallel scheme remains unknown.

The problem belongs to a main research theme in multi-parameter quantum estimation, which is to quantify the minimal tradeoff among the precisions of estimating multiple parameters\cite{vidrighin2014,crowley2014,Yuan2016PRL,Ragy2016,Chen_2017,Yue2014,Zhang2014, ChenHZ2019,Gill2000,Bagan2006,LiNan2016,Zhu2018,Magdalena2016,Albarelli2019,Francesco2019,Liu_2019}. Over the past decades there have been extensive studies on this theme, however, the minimal tradeoff remains only known for very limited cases\cite{vidrighin2014,crowley2014,Yuan2016PRL,Ragy2016}. The study on the tradeoff induced by the incompatibility of the measurements has made much progress\cite{HOLE82,Barndorff_Nielsen_2000,Gill2000,Bagan2006,LiNan2016,Zhu2018,Magdalena2016,PhysRevLett.119.130504,Yang2019,Carollo_2019,Francesco2019,Liu_2019,Francesco20192,Albarelli2019,Tsang2019}. However, the tradeoff induced by the incompatibility of the optimal probe states is much less understood. We present an approach to study the tradeoff induced by the incompatibility of the optimal probe states and obtain the minimal tradeoff for the multi-parameter quantum magmetometry under the parallel scheme. Here the figure of merit can be taken as the sum of arbitrarily weighted variance and the generated unitary does not need to be close to the Identity operator. The obtained precision not only provides a fundamental limit for multi-parameter quantum magnetometry under the parallel scheme, but can also be used to calibrate the ultimate performances of the quantum reference frame alignment, quantum gyroscope, etc. We note that additional controls during the evolution are not included in the parallel scheme. Controlled schemes on small systems have been studied in \cite{Yuan2016PRL,Jing2017}. Accurate controls on systems with large $N$ are typically hard to implement, and in some settings, such as in quantum reference frame alignment, the information is encoded in the unitary which can not be altered with controls.

We first use spin-1/2 as the probe for the estimation of the magnetic field. The dynamics for a spin-1/2 in a magnetic field can be described by the Hamiltonian $H=\vec{B}\cdot \vec{\sigma}=B_1\sigma_1+B_2\sigma_2+B_3\sigma_3$, where $\sigma_1=\left(
\begin{array}{cc}
0 & 1 \\
1 & 0 \\
\end{array}
\right)$,
$\sigma_2=\left(
\begin{array}{cc}
0 & -\rmi \\
\rmi & 0 \\
\end{array}
\right)$, and
$\sigma_3=\left(
\begin{array}{cc}
1 & 0 \\
0 & -1 \\
\end{array}
\right)$
are the Pauli matrices. This can be equivalently written as $H=B\vec{n}\cdot \vec{\sigma}$ with $B=\sqrt{B_1^2+B_2^2+B_3^2}$ as the magnitude of the magnetic field and $\vec{n}=(\sin\theta\cos\phi,\sin\theta\sin\phi,\cos\theta)$ as the direction of the magnetic field. After an evolution time $t$, the dynamics generates a $SU(2)$ operator as
$U_s
=e^{-\rmi\alpha\vec{n}\cdot\vec{\sigma}}$ with $\alpha=Bt$. As we allow the figure of merit taken as the sum of arbitrarily weighted variance, we can just consider the precision for the simultaneous estimation of $\alpha, \theta$ and $\phi$ with the figure of merit as $w_1\delta\hat{\alpha}^2+w_2\delta\hat{\theta}^2+w_3\delta\hat{\phi}^2$, here $w_1, w_2, w_3> 0$ are the weights and $\delta\hat{x}^2=E[(\hat{x}-x)^2]$ denotes the variance for an unbiased estimator of a parameter. The estimation for various other parameters can be expressed in terms of $\alpha$, $\theta$ and $\phi$ with different weights. For example, the precision for the estimation of $B$ is related to $\alpha$ as $\delta \hat{B}^2=\frac{\delta \hat{\alpha}^2}{t^2}$, thus the sum of equally weighted variance for $(B, \theta, \phi)$ can be written as  $\delta\hat{B}^2+\delta\hat{\theta}^2+\delta\hat{\phi}^2=\frac{1}{t^2}\delta\hat{\alpha}^2+\delta\hat{\theta}^2+\delta\hat{\phi}^2$. Similarly the sum of equally weighted variance for the estimation of $(B_1, B_2, B_3)$ can be expressed as $\delta\hat{B}_1^2+\delta\hat{B}_2^2+\delta\hat{B}_3^2=\frac{1}{t^2}(\delta\hat{\alpha}^2+\alpha^2\delta\hat{\theta}^2+\alpha^2\sin^2\theta\delta\hat{\phi}^2)$. This differs from most previous studies which take the figure of merit as the sum of equally weighted variance under a specific parametrizations\cite{Kahn2007,Manuel2004,Fujiwara2001,Imai2007,Manuel20042,Manuel2005,Datta2016,Liu_2017}.

The precision limit for the estimation of a parameter, involving $m$ repetitions of the experiment(here the repetition, $m$, represents the classical effect, which we will neglect in the rest of the article), is given by the quantum Cr\'amer-Rao bound (QCRB)
\begin{equation}
\label{eq:uncertainty relation}
\delta \hat{x}^2\geq\frac{1}{mJ_x}=\frac{1}{4m\<\Delta H_x^2\>},
\end{equation}
here $H_x$ is the generator of the corresponding parameter $x$\cite{HELS67,HOLE82,BRAU96}, which is defined as $H_x\equiv\rmi U_s^\dagger(\partial_x U_s)$, $U_s=e^{-\rmi\alpha\vec{n}\cdot\vec{\sigma}}$ is the generated unitary\cite{Wilcox1967,Brody_2013,pang2014,Liu2015,Liu_2019}, $\<\Delta H_x^2\>=\bra{\Psi}H_x^2\ket{\Psi}-\bra{\Psi}H_x\ket{\Psi}^2$ is the variance of $H_x$ with respect to the initial probe state $|\Psi\rangle$, $J_x=4\<\Delta H_x^2\>$ is the quantum Fisher information. For $x\in \{\alpha, \theta, \phi\}$, the corresponding generator can be obtained as
\begin{eqnarray}
\aligned
H_{\alpha}&=c_{\alpha}\vec{n}_{\alpha}\cdot\vec{\sigma},\\
H_{\theta}&=c_{\theta}\vec{n}_{\theta}\cdot\vec{\sigma},\\
H_{\phi}&=c_{\phi}\vec{n}_{\phi}\cdot\vec{\sigma},
\endaligned
\end{eqnarray}
with $c_\alpha=1$, $\vec{n}_\alpha=\vec{n}=(\sin\theta\cos\phi,\sin\theta\sin\phi,\cos\theta)$, $c_\theta=\sin\alpha$, $\vec{n}_\theta= \cos\alpha\vec{n}_1-\sin\alpha\vec{n}_2$, $c_\phi=\sin\alpha\sin\theta$ and $\vec{n}_\phi=\cos\alpha\vec{n}_2+\sin\alpha\vec{n}_1$ respectively,
here $\vec{n}_1=\partial_\theta\vec{n}=(\cos\theta\cos\phi,\cos\theta\sin\phi,-\sin\theta)$, $\vec{n}_2=\vec{n}\times\vec{n}_1=(-\sin\phi,\cos\phi,0)$. $\vec{n}$, $\vec{n}_1$ and $\vec{n}_2$ are orthogonal to each other.

With $N$ spins interacting with the field and a possible ancilla, the generator for each parameter is
$H_x^{(N)}=\sum\limits_{k=0}^{N-1} H_x^{[k]},$
where $H_x^{[k]}=I\otimes \cdots \otimes I\otimes H_x\otimes I\cdots \otimes I\otimes I_A$ denotes the generator on the $k$th spin, $I$ denotes the Identity operator and $I_A$ denotes the Identity operator on the ancilla. The variance of $H_x^{(N)}$ is given by
\begin{equation}\label{eq:Nvariance}
\left\<\Delta \left[H_x^{(N)}\right]^2\right\>= \left\<\left(H_x^{(N)}\right)^2\right\>-\left\<H_x^{(N)}\right\>^2,
\end{equation}
where the first term can be expanded as
\begin{equation}\label{eq:first}
\aligned
\left\<\left(H_x^{(N)}\right)^2\right\>&=\sum\limits_{k=0}^{N-1} \left\<\left(H_x^{[k]}\right)^2\right\>+\sum\limits_{j\neq k} \left\<H_x^{[j]}H_x^{[k]}\right\>\\
&=c_x^2\left(N+\sum\limits_{j\neq k}r^{(j,k)}_{xx}\right),
\endaligned
\end{equation}
and the second term as
$\left\<H_x^{(N)}\right\>^2=c_x^2\left(\sum\limits_{k=0}^{N-1} r_x^{(k)}\right)^2$,
here we denote $r^{(j,k)}_{xx}=\tr\left[\rho^{(j,k)}\left(\vec{n}_x\cdot\vec{\sigma}\otimes\vec{n}_x\cdot\vec{\sigma}\right)\right]\leq 1$, $r^{(k)}_{x}=\tr\left(\rho^{(k)}\vec{n}_x\cdot\vec{\sigma}\right)$ with $\rho^{(j,k)}$ as the reduced density matrix for the $j$-th and $k$-th spin and $\rho^{(k)}$ as the reduced density matrix for the $k$-th spin\cite{Manuel2005,Datta2016, Liu_2017}. It is easy to see the same formula holds without the ancillary system (which corresponds to taking $I_A=1$), however, the ancillary system provides more room on the choices of $\rho^{(j,k)}$, which can be seen in the analysis of the optimal states below.

It can be seen that $\left\<\Delta \left[H_x^{(N)}\right]^2\right\> \leq N^2c_x^2$, where the equality can be reached iff $\sum\limits_{k=0}^{N-1} r_x^{(k)}=0$ and $r^{(j,k)}_{xx}=1$ for all $j, k$. For a single parameter, this upper bound, which corresponds to the highest precision achievable for the estimation of the corresponding parameter, can be saturated by choosing the probe state as the GHZ-type state, $\ket{\Phi_x}=\frac1{\sqrt{2}}\left(\ket{+_x}^{\otimes N}+\ket{-_x}^{\otimes N}\right)$, where $\ket{\pm_x}$ are the eigenstates of $H_x$. It is easy to check that the reduced two-spin state is $\rho^{(j,k)}=\frac1{2}\left(\ketbra{+_x+_x}{+_x+_x}+\ketbra{-_x-_x}{-_x-_x}\right)=\frac{1}{4}(I^{(j,k)}+\vec{n}_x\cdot\vec{\sigma}\otimes\vec{n}_x\cdot\vec{\sigma})$ for all $(j,k)$ and the reduced single spin state is $\rho^{(k)}=\frac{I^{(k)}}{2}$ for all $k$, thus $r^{(j,k)}_{xx}=1$ and $r_x^{(k)}=0$. The highest precision for a single parameter can thus be achieved.

For the estimation of multiple parameters, however, the issue is much more complicated.
A main research theme in multi-parameter quantum estimation is to clarify whether it is possible to achieve the highest precision for all parameters simultaneously and calibrate the minimal tradeoff among the precisions of different parameters when it is not possible.

To calibrate the minimal tradeoff, we write a general two-qubit state as
\begin{eqnarray}\label{eq:two}
\aligned
\rho^{(j,k)}=\frac{1}{4}[&I^{(j,k)}+\sum_{l}
r_l^{(j)}\sigma_l^{(j)}\otimes I^{(k)}+\sum_{p}r_p^{(k)}I^{(j)}\otimes \sigma_p^{(k)}\\
&+\sum_{l,p}r_{lp}^{(j,k)}\sigma_l^{(j)}\otimes\sigma_p^{(k)}],
\endaligned
\end{eqnarray}
here $l,p\in\{\alpha,\theta,\phi\}$, and we have denoted $\sigma_{\alpha}=\vec{n}_{\alpha}\cdot\vec{\sigma}$, $\sigma_{\theta}=\vec{n}_{\theta}\cdot\vec{\sigma}$ and $\sigma_{\phi}=\vec{n}_{\phi}\cdot\vec{\sigma}$. Now let $U=e^{i\frac{\alpha}{2} \vec{n}\cdot \vec{\sigma}}e^{-i\frac{\phi}{2}\sigma_3}e^{-i\frac{\theta}{2}\sigma_2}$, which is the unitary that satisfies $U\sigma_1U^\dagger=\sigma_{\theta}$,  $U\sigma_2U^\dagger=\sigma_{\phi}$ and  $U\sigma_3U^\dagger=\sigma_{\alpha}$, and let $\ket{\Psi^{(j,k)}_-}=\frac{U}{\sqrt{2}}(\ket{01}-\ket{10})$, then $\ket{\Psi^{(j,k)}_-}\bra{\Psi^{(j,k)}_-}=\frac{1}{4}[I^{(j,k)}-\sum_{x\in\{\alpha,\theta,\phi\}} \sigma_x^{(j)}\otimes\sigma_x^{(k)}]$.
As $\rho^{(j,k)}\geq 0$, we have $\bra{\Psi^{(j,k)}_-}\rho^{(j,k)}\ket{\Psi^{(j,k)}_-}=\tr(\rho^{(j,k)}\ket{\Psi^{(j,k)}_-}\bra{\Psi^{(j,k)}_-})\geq 0$, which gives a constraint as $r_{\alpha\alpha}^{(j,k)}+r_{\theta\theta}^{(j,k)}+r_{\phi\phi}^{(j,k)}\leq 1$. This clearly shows that $r_{\alpha\alpha}^{(j,k)}$, $r_{\theta\theta}^{(j,k)}$, $r_{\phi\phi}^{(j,k)}$ can not equal to $1$ simultaneously and the tradeoff among the precisions of different parameters is unavoidable. It turns out such constraint fully calibrates the minimal tradeoff among the precisions.


We consider the figure of merit as $w_{\alpha}\delta \hat{\alpha}^2+w_{\theta}\delta \hat{\theta}^2+w_{\phi}\delta \hat{\phi}^2$, where $w_i>0$ are weights that can be chosen arbitrarily according to specific needs. Under the constraint $r_{\alpha\alpha}^{(j,k)}+r_{\theta\theta}^{(j,k)}+r_{\phi\phi}^{(j,k)}\leq 1$, the sum of weighted variance is bounded below as(see supplemental material for derivation)
\begin{equation}\label{eq:bound}
w_{\alpha}\delta \hat{\alpha}^2+w_{\theta}\delta \hat{\theta}^2+w_{\phi}\delta \hat{\phi}^2\geq \frac{(\sqrt{w_{\alpha}}+\frac{\sqrt{w_{\theta}}}{|\sin\alpha|}+\frac{\sqrt{w_{\phi}}}{|\sin\alpha\sin\theta|})^2}{4N(N+2)}.
\end{equation}
The lower bound can be saturated when the reduced two-qubit state takes the form as $\rho^{(j,k)}=\frac{1}{4}[I^{(j,k)}+ \tilde{r}_{\alpha\alpha}\sigma_\alpha^{(j)}\otimes\sigma_\alpha^{(k)}+\tilde{r}_{\theta\theta}\sigma_\theta^{(j)}\otimes\sigma_\theta^{(k)}+\tilde{r}_{\phi\phi}\sigma_\phi^{(j)}\otimes\sigma_\phi^{(k)}]$ for all $0\leq j<k\leq N-1$ with
\begin{eqnarray}\label{eq:optimal}
\aligned
\tilde{r}_{\alpha\alpha}&=\frac{(N+1)\sqrt{w_{\alpha}}-\frac{\sqrt{w_{\theta}}}{|\sin\alpha|}-\frac{\sqrt{w_{\phi}}}{|\sin\alpha\sin\theta|}}{(N-1)(\sqrt{w_{\alpha}}+\frac{\sqrt{w_{\theta}}}{|\sin\alpha|}+\frac{\sqrt{w_{\phi}}}{|\sin\alpha\sin\theta|})},\\ \tilde{r}_{\theta\theta}&=\frac{(N+1)\frac{\sqrt{w_{\theta}}}{|\sin\alpha|}-\sqrt{w_{\alpha}}-\frac{\sqrt{w_{\phi}}}{|\sin\alpha\sin\theta|}}{(N-1)(\sqrt{w_{\alpha}}+\frac{\sqrt{w_{\theta}}}{|\sin\alpha|}+\frac{\sqrt{w_{\phi}}}{|\sin\alpha\sin\theta|})},\\ \tilde{r}_{\phi\phi}&=\frac{(N+1)\frac{\sqrt{w_{\phi}}}{|\sin\alpha\sin\theta|}-\sqrt{w_{\alpha}}-\frac{\sqrt{w_{\theta}}}{|\sin\alpha|}}{(N-1)(\sqrt{w_{\alpha}}+\frac{\sqrt{w_{\theta}}}{|\sin\alpha|}+\frac{\sqrt{w_{\phi}}}{|\sin\alpha\sin\theta|})}. \endaligned
\end{eqnarray}
The problem now is to identify the states whose reduced two-spin states are of this form which leads to the minimal tradeoff among the precisions.

By employing a qutrit(or three levels in two additional spin-1/2) as the ancillary system we can prepare the probe state as
\begin{eqnarray}\label{eq:optimalstate}
|\Phi_o\rangle=s_\alpha|\Phi_{\alpha}\rangle\otimes|0\rangle+s_\theta|\Phi_{\theta}\rangle\otimes|1\rangle+s_\phi|\Phi_{\phi}\rangle\otimes|2\rangle,
\end{eqnarray}
here $|\Phi_{x}\rangle=\frac{1}{\sqrt{2}}(|+_x\rangle^{\otimes N}+|-_x\rangle^{\otimes N})$ with $|\pm_x\rangle$ as the eigen-states of $H_x$, $x\in \{\alpha,\theta,\phi\}$, $N$ is the number of spins that interact with the magnetic field.
The reduced two-spin state of this state is
\begin{eqnarray}\label{eq:reducedAncilla}
\aligned
\rho^{(j,k)}=\frac{1}{4}[&I^{(j,k)}+|s_\alpha|^2\sigma_\alpha^{(j)}\otimes\sigma_\alpha^{(k)}\\
+&|s_\theta|^2\sigma_\theta^{(j)}\otimes\sigma_\theta^{(k)}+|s_\phi|^2\sigma_\phi^{(j)}\otimes\sigma_\phi^{(k)}]
\endaligned
\end{eqnarray}
for all $0\leq j<k\leq N-1$ and the reduced single spin state is $\rho^{(k)}=\frac{I^{(k)}}{2}$ for all $0\le k\le N-1$.
For multiple parameters the QCRB is given by $Cov(\hat{x})\geq J^{-1}$, where $J$ is now the quantum Fisher information matrix whose entries can be obtained from the generators as
$   J_{lp}=4[\frac{1}{2}\bra{\Psi}\{H_{l}^{(N)},H_{p}^{(N)}\}\ket{\Psi}-\bra{\Psi}H_{l}^{(N)}\ket{\Psi}\bra{\Psi}H_{p}^{(N)}\ket{\Psi}]$,
$l,p\in\{\alpha,\theta,\phi\}$. For the state in Eq.(\ref{eq:reducedAncilla}), it is straightforward to check (see supplement) that $J$ is a diagonal matrix.
If the optimal $\tilde{r}_{\alpha\alpha}$, $\tilde{r}_{\theta\theta}$ and $\tilde{r}_{\phi\phi}$ in Eq.(\ref{eq:optimal}) are all non-negative, then by choosing $s_\alpha=\sqrt{\tilde{r}_{\alpha\alpha}}$, $s_\theta=\sqrt{\tilde{r}_{\theta\theta}}$ and $s_\phi=\sqrt{\tilde{r}_{\phi\phi}}$, the ultimate lower bound in Eq.(\ref{eq:bound}) is saturated. For sufficiently large $N$, this is always the case. It is also straightforward to check the weak commutativity condition, $\left<\Psi(\alpha,\theta,\psi)|[L_l,L_p]|\Psi(\alpha,\theta,\psi)\right>=0$, holds for all $l,p\in\{\alpha,\theta,\phi\}$\cite{Matsumoto2002}, here $L_p$ is the symmetric logarithmic derivatives(SLD) for parameter $p\in \{\alpha, \theta,\phi\}$, which is the solution to the equation $\partial_p \rho=\frac{1}{2}(L_p\rho+\rho L_p)$. This condition ensures the existence of a measurement saturating the QCRB\cite{Matsumoto2002,Ragy2016,Yang2019}. The condition can be simplified as $\mathrm{Im}[\left<\partial_l\Psi(\alpha,\theta,\psi)|\partial_p\Psi(\alpha,\theta,\psi)\right>]=0$. For pure states, this ensures that among all possible SLDs (note for pure states the SLD for a parameter is not unique) there exists a set of commuting SLDs\cite{Matsumoto2002,Yang2019}.
We provide an explicit construction of the optimal measurement in the supplemental material. The lower bound in Eq.(\ref{eq:bound}) can thus always be achieved for sufficiently large $N$, which is the ultimate precision limit that can be achieved under the paralell scheme.

If some $\tilde{r}_{xx}$ in Eq.(\ref{eq:optimal}), $x\in \{\alpha,\theta,\phi\}$, are negative for small $N$, then the lower bound in Eq.(\ref{eq:bound}) can not be saturated by the probe states of this form. The best precision achieved by these states can be obtained by optimizing the coefficients $s_\alpha$, $s_\theta$ and $s_\phi$, which can be analytically obtained(see supplementary material). In Fig.~\ref{fig:scheme2} we plotted the precisions that can be achieved for different weights and $N$, it can be seen that the obtained precision is already close to the ultimate bound even for small $N$, and it saturates the ultimate bound when $N$ gets large.

\begin{figure}[t]
	\center{\includegraphics[width=0.45\textwidth]{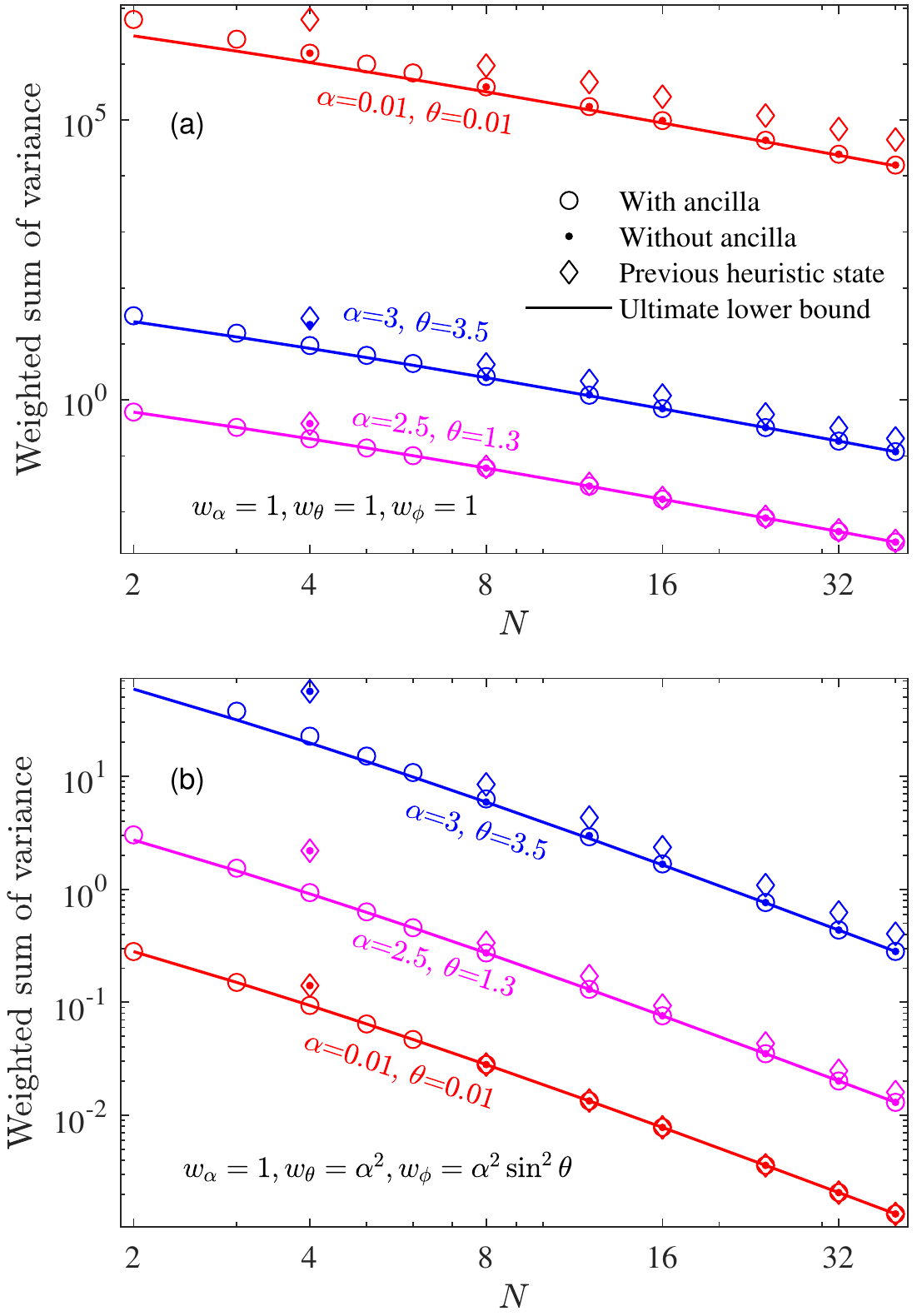}}
	\caption{\label{fig:scheme2} 
(a) Weighted sum of variance with $w_\alpha=1$, $w_\theta=1$ and $w_\phi=1$, which corresponds to $\delta \hat{\alpha}^2+\delta \hat{\theta}^2+\delta \hat{\phi}^2$.	(b) Weighted sum of variance with $w_\alpha=1, w_\theta=\alpha^2$ and $w_\phi=\alpha^2\sin^2\theta$, which corresponds to $\delta \hat{B}_1^2+\delta \hat{B}_2^2+\delta \hat{B}_3^2$. Three typical sets of values as specified in the figure for each case. The time has been normalized, i.e., t=1.}
\end{figure}

The ultimate lower bound in Eq.(\ref{eq:bound}) can also be achieved without the ancillary system when $N\rightarrow \infty$ by preparing the probe state as $|\Phi_o\rangle=\sqrt{\tilde{r}_{\alpha\alpha}}|\Phi_{\alpha}\rangle+\sqrt{\tilde{r}_{\theta\theta}}|\Phi_{\theta}\rangle+\sqrt{\tilde{r}_{\phi\phi}}|\Phi_{\phi}\rangle$.
The heuristic state given in \cite{Datta2016} also takes this form but with equal coefficients.
We note that without the ancillary system, the ultimate lower bound in Eq.(\ref{eq:bound}) can only be achieved when $N\rightarrow \infty$, even all $\tilde{r}_{xx}$ in Eq.(\ref{eq:optimal}) are non-negative(see the analysis for finite $N$ in supplement). While with the ancillary system, the ultimate lower bound can be exactly saturated as long as all $\tilde{r}_{xx}$ in Eq.(\ref{eq:bound}) are non-negative. 

We now compare the obtained precision with previous results.
For the estimation of the three components of the magnetic field, $B_1=\frac{\alpha}{t} \sin\theta\cos\phi$, $B_2=\frac{\alpha}{t} \sin\theta\sin\phi$, $B_3=\frac{\alpha}{t}\cos\theta$, we have
$\delta \hat{B}_1^2+\delta \hat{B}_2^2+\delta \hat{B}_3^2
=\frac{\delta \hat{\alpha}^2+\alpha^2\delta \hat{\theta}^2+\alpha^2\sin^2\theta\delta\hat{\phi}^2}{t^2}$,
which corresponds to taking $w_{\alpha}=\frac{1}{t^2}$, $w_{\theta}=\frac{\alpha^2}{t^2}$, $w_{\phi}=\frac{\alpha^2\sin^2\theta}{t^2}$. The ultimate precision is then given by
\begin{equation}\label{eq:x}
\delta \hat{B}_1^2+\delta \hat{B}_2^2+\delta \hat{B}_3^2\geq \frac{(1+2|\frac{\alpha}{\sin\alpha}|)^2}{4N(N+2)t^2}.
\end{equation}
While the best precision obtained previously with the heuristic state\cite{Datta2016} is
$\delta \hat{B}_1^2+\delta \hat{B}_2^2+\delta \hat{B}_3^2\geq \frac{3(1+2\frac{\alpha^2}{\sin^2\alpha})}{4N(N+2)t^2}.$
They are equivalent only at the weak limit when $\alpha=Bt\rightarrow 0$, as shown in Fig.~\ref{fig:scheme2}(b). The difference between them, which is $\frac{2(|\frac{\alpha}{\sin\alpha}|-1)^2}{4N(N+2)t^2}$, can be large particularly when $\alpha\rightarrow m\pi$.

This approach can be generalized to general spin-$S$, where the Hamiltonian is $H=B\vec{n}\cdot \vec{S}=B_1S_1+B_2S_2+B_3S_3$, here $\vec{n}=(\sin\theta\cos\phi, \sin\theta\sin\phi,\cos\theta)$, $S_1$, $S_2$ and $S_3$ are general spin operators which satisfy $[S_1,S_2]=iS_3$, $[S_2,S_3]=iS_1$, $[S_3,S_1]=iS_2$(with this commutation relation, $S_i=\frac{\sigma_i}{2}$ when $S=1/2$, which has an extra factor of $\frac{1}{2}$ comparing with the Pauli matrices). For $S>1/2$, $S_i^2\not\propto I$, but $S_1^2+S_2^2+S_3^2=S(S+1)I$. The generators for $\alpha$, $\theta$ and $\phi$ can be similarly obtained as $H_\alpha=c_\alpha{S_\alpha}$, $H_\theta=c_\theta S_\theta$ and $H_\phi=c_\phi S_\phi$, where $c_\alpha=1$, $S_\alpha=\bm{n}_\alpha\cdot\bm{S}$, $\bm{n}_\alpha=\bm{n}=(\sin\theta\cos\phi,\sin\theta\sin\phi,\cos\theta)$, $c_\theta=2\sin\frac{\alpha}{2}$, $S_\theta=\bm{n}_\theta\cdot\bm{S}$, $\bm{n}_\theta=\cos\frac{\alpha}{2}\bm{n}_1-\sin\frac{\alpha}{2}\bm{n}_2$, $c_\phi=2\sin\frac{\alpha}{2}\sin\theta$, $S_\phi=\bm{n}_\phi\cdot\bm{S}$, $\bm{n}_\phi=\sin\frac{\alpha}{2}\bm{n}_1+\cos\frac{\alpha}{2}\bm{n}_2$ and
$\bm{n_1}=\partial_\theta\bm{n}$, $\bm{n_2}=\bm{n}\times\bm{n_1}$. We can obtain the lower bound as
\begin{equation}\label{eq:S}
  \begin{aligned}
   & w_\alpha\delta \hat{\alpha}^{2}+w_\theta\delta \hat{\theta}^{2}+w_\phi\delta \hat{\phi}^{2}\\
    &\geq \frac{1}{4}\frac{\left(\sqrt{w_{\alpha}}+\frac{\sqrt{w_{\theta}}}{|2\sin\frac{\alpha}{2}|}+\frac{\sqrt{w_{\phi}}}{|2\sin\frac{\alpha}{2}\sin\theta|}\right)^2} {\left[ \sum_{k=0}^{N-1}\sum_{x\in\{\alpha,\theta,\phi\}} r^{(k)}_{xx}+\sum_{j\neq k}\sum_{x\in\{\alpha,\theta,\phi\}} r^{(j,k)}_{xx} \right]},
  \end{aligned}
\end{equation}
here $r_{xx}^{(k)}=\tr(\rho^{(k)}S_x^2)$, $r_{xx}^{(j,k)}=\tr(\rho^{(j,k)}S_x\otimes S_x)$, $\forall x\in\{\alpha,\theta,\phi\}$. It is easy to get $\sum_{x\in\{\alpha,\theta,\phi\}}r_{xx}^{(k)}=\tr(\rho^{(k)}\sum_{x\in\{\alpha,\theta,\phi\}}S_x^2)=S(S+1)$. The constrains on $r_{xx}^{(j,k)}$, however, are much harder to obtain for $S>1/2$. In the supplement material, we show that $\sum_{x\in\{\alpha,\theta,\phi\}}S_x\otimes S_x\leq S^2 I$, thus $\sum_{x\in\{\alpha,\theta,\phi\}}r_{xx}^{(j,k)}\leq S^2$. The ultimate lower bound is then
\begin{equation}\label{eq:Sbound}
  \begin{aligned}
   & w_\alpha\delta \hat{\alpha}^{2}+w_\theta\delta \hat{\theta}^{2}+w_\phi\delta \hat{\phi}^{2}\\
    &\geq \frac{\left(\sqrt{w_{\alpha}}+\frac{\sqrt{w_{\theta}}}{|2\sin\frac{\alpha}{2}|}+\frac{\sqrt{w_{\phi}}}{|2\sin\frac{\alpha}{2}\sin\theta|}\right)^2} {4NS(NS+1)}.
  \end{aligned}
\end{equation}
With an ancillary qutrit, the ultimate lower bound can be saturated for sufficiently large $NS$ with the state
$|\Phi_o\rangle=s_\alpha|\Phi_{\alpha}\rangle\otimes|0\rangle+s_\theta|\Phi_{\theta}\rangle\otimes|1\rangle+s_\phi|\Phi_{\phi}\rangle\otimes|2\rangle$,
here $|\Phi_{x}\rangle=\frac{1}{\sqrt{2}}(|+_x\rangle^{\otimes N}+|-_x\rangle^{\otimes N})$ with $|\pm_x\rangle$ as the eigen-states of $S_x$ corresponding to the eigenvalue $\pm S$ respectively, $\forall x\in \{\alpha,\theta,\phi\}$, and the coefficients should satisfy
\begin{equation}
  \begin{aligned}
    |s_\alpha|^2&=\frac{(2NS+1)\sqrt{w_{\alpha}}-\frac{\sqrt{w_{\theta}}}{2|\sin\frac{\alpha}{2}|}-\frac{\sqrt{w_{\phi}}}{2|\sin\frac{\alpha}{2} \sin \theta|}}{(2NS-1)(\sqrt{w_{\alpha}}+\frac{\sqrt{w_{\theta}}}{2|\sin\frac{\alpha}{2}|}+\frac{\sqrt{w_{\phi}}}{2|\sin\frac{\alpha}{2} \sin \theta|})},\\
    |s_\theta|^2&=\frac{(2NS+1)\frac{\sqrt{w_{\theta}}}{2|\sin\frac{\alpha}{2}|}-\frac{\sqrt{w_{\phi}}}{2|\sin\frac{\alpha}{2} \sin \theta|}-\sqrt{w_{\alpha}}}{(2NS-1)(\sqrt{w_{\alpha}}+\frac{\sqrt{w_{\theta}}}{2|\sin\frac{\alpha}{2}|}+\frac{\sqrt{w_{\phi}}}{2|\sin\frac{\alpha}{2} \sin \theta|})},\\
    |s_\phi|^2&=\frac{(2NS+1)\frac{\sqrt{w_{\phi}}}{2|\sin\frac{\alpha}{2} \sin \theta|}-\sqrt{w_{\alpha}}-\frac{\sqrt{w_{\theta}}}{2|\sin\frac{\alpha}{2}|}}{(2NS-1)(\sqrt{w_{\alpha}}+\frac{\sqrt{w_{\theta}}}{2|\sin\frac{\alpha}{2}|}+\frac{\sqrt{w_{\phi}}}{2|\sin\frac{\alpha}{2} \sin \theta|})},
  \end{aligned}
\end{equation}
which always have solutions when $NS$ is sufficiently large. It is also straightforward to check that the weak commutativity condition also holds, the ultimate lower bound can thus always be saturated for sufficiently large $N$ or $S$.


\emph{Summary} The obtained precision is the ultimate precision that can be achieved under the parallel scheme, which is of fundamental interest and importance in quantum metrology. It can also be directly used to calibrate the performance of quantum gyroscope and quantum reference frame alignment. Our approach connects the tradeoff directly to the constraints on the probe states and the generators, which makes the tradeoff more transparent. We expect it can lead to many nontrivial (may not be always achievable, nevertheless useful) bounds in various other scenarios. Future studies can include measurements suitable for specific physical settings and generalization to noisy dynamics via the purification approach\cite{Fujiwara2008,Escher2011,Rafal2012,yuan2017quantum}

\begin{acknowledgments}
The work at USTC is supported by the National Natural Science Foundation of China under Grants (Nos. 61905234, 11974335, 11574291 and 11774334),  the National Key Research and Development Program of China (Grant No.2018YFA0306400 and 2017YFA0304100),  Key Research Program of Frontier Sciences, CAS (No.QYZDY-SSW-SLH003), Anhui Initiative in Quantum Information Technologies. The work at CUHK is supported by the Research Grants Council of Hong Kong(GRF No. 14207717).
\end{acknowledgments}

\appendix
\begin{widetext}
\section{Ultimate lower bound on arbitrarily weighted sum of variance}
The precision limit for the estimation of a single parameter is given by
\begin{equation}
\label{eq:uncertainty relationapp}
\delta \hat{x}^2\geq\frac{1}{4\<\Delta H_x^2\>},
\end{equation}
here $H_x$ is the generator corresponding to the parameter $x\in\{\alpha,\theta,\phi\}$.

With $N$ copies of the operator acting on $N$ spins, the generator for each parameter is
\begin{equation}\label{eq:N copies parallel}
H_x^{(N)}=\sum\limits_{k=0}^{N-1} H_x^{[k]},
\end{equation}
where $H_x^{[k]}=I\otimes \cdots \otimes I\otimes H_x\otimes I\cdots \otimes I\otimes I_A$ denotes the generator on the $k$th spin. The variance of $H_x^{(N)}$ is given by
\begin{equation}\label{eq:N copies variance}
\<\Delta \left[H_x^{(N)}\right]^2\>= \left\<\left(H_x^{(N)}\right)^2\right\>-\left\<H_x^{(N)}\right\>^2,
\end{equation}
where the first term can be expanded as
\begin{equation}\label{eq:first term}
\aligned
\left\<\left(H_x^{(N)}\right)^2\right\>&=\sum\limits_{k=0}^{N-1} \left\<\left(H_x^{[k]}\right)^2\right\>+\sum\limits_{j\neq k} \left\<H_x^{(j)}H_x^{[k]}\right\>\\
&=c_x^2\left(N+\sum\limits_{j\neq k}r^{(j,k)}_{xx}\right),
\endaligned
\end{equation}
and the second term as
\begin{equation}\label{eq:second term}
\left\<H_x^{(N)}\right\>^2=c_x^2\left(\sum\limits_{k=0}^{N-1} r_x^{(k)}\right)^2,
\end{equation}
here $c_\alpha=1$, $\vec{n}_\alpha=\vec{n}=(\sin\theta\cos\phi,\sin\theta\sin\phi,\cos\theta)$, $c_\theta=\sin\alpha$, $\vec{n}_\theta= \cos\alpha\vec{n}_1+\sin\alpha\vec{n}_2$, $c_\phi=\sin\alpha\sin\theta$, $\vec{n}_\phi=\cos\alpha\vec{n}_2-\sin\alpha\vec{n}_1$,  $r^{(j,k)}_{xx}=\tr\left[\rho^{(j,k)}\left(\vec{n}_x\cdot\vec{\sigma}\otimes\vec{n}_x\cdot\vec{\sigma}\right)\right]\leq 1$, $r^{(k)}_{x}=\tr\left(\rho^{(k)}\vec{n}_x\cdot\vec{\sigma}\right)$ with $\rho^{(j,k)}$ as the reduced density matrix for the $j$-th and $k$-th spin and $\rho^{(k)}$ as the reduced density matrix for the $k$-th spin.

For each parameter $x\in\{\alpha,\theta,\phi\}$, we have $\delta \hat{x}^2\geq \frac{1}{4 \<\Delta[H_x^{(N)}]^2\>}$, thus
\begin{eqnarray}
\aligned
&w_{\alpha}\delta \hat{\alpha}^2+w_{\theta}\delta \hat{\theta}^2+w_{\phi}\delta \hat{\phi}^2\\
\geq &\frac{1}{4}\left( \frac{w_{\alpha}}{\<\Delta \left[H_{\alpha}^{(N)}\right]^2\>}+\frac{w_{\theta}}{\<\Delta \left[H_{\theta}^{(N)}\right]^2\>}+\frac{w_{\phi}}{\<\Delta \left[H_{\phi}^{(N)}\right]^2\>}\right)\\
=&\frac{1}{4}\left( \frac{w_{\alpha}}{\left(N+\sum\limits_{j\neq k}r^{(j,k)}_{\alpha\alpha}\right)-\left(\sum\limits_{k=0}^{N-1} r_\alpha^{(k)}\right)^2}+\frac{w_{\theta}/\sin^2\alpha}{\left(N+\sum\limits_{j\neq k}r^{(j,k)}_{\theta\theta}\right)-\left(\sum\limits_{k=0}^{N-1} r_\theta^{(k)}\right)^2}+\frac{w_{\phi}/\sin^2\alpha\sin^2\theta}{\left(N+\sum\limits_{j\neq k}r^{(j,k)}_{\phi\phi}\right)-\left(\sum\limits_{k=0}^{N-1}r_\phi^{(k)}\right)^2}\right)\\
\geq&\frac{1}{4}\left( \frac{w_{\alpha}}{\left(N+\sum\limits_{j\neq k}r^{(j,k)}_{\alpha\alpha}\right)}+\frac{w_{\theta}/\sin^2\alpha}{\left(N+\sum\limits_{j\neq k}r^{(j,k)}_{\theta\theta}\right)}+\frac{w_{\phi}/\sin^2\alpha\sin^2\theta}{\left(N+\sum\limits_{j\neq k}r^{(j,k)}_{\phi\phi}\right)}\right).
\endaligned
\end{eqnarray}
From Cauchy-Schwarz inequality, we get
\begin{eqnarray}
\aligned
&\left( \frac{w_{\alpha}}{\left(N+\sum\limits_{j\neq k}r^{(j,k)}_{\alpha\alpha}\right)}+\frac{w_{\theta}/\sin^2\alpha}{\left(N+\sum\limits_{j\neq k}r^{(j,k)}_{\theta\theta}\right)}+\frac{w_{\phi}/\sin^2\alpha\sin^2\theta}{\left(N+\sum\limits_{j\neq k}r^{(j,k)}_{\phi\phi}\right)}\right)\left(N+\sum\limits_{j\neq k}r^{(j,k)}_{\alpha\alpha}+N+\sum\limits_{j\neq k}r^{(j,k)}_{\theta\theta}+N+\sum\limits_{j\neq k}r^{(j,k)}_{\phi\phi}\right)\\
\geq &(\sqrt{w_{\alpha}}+\frac{\sqrt{w_{\theta}}}{|\sin\alpha|}|+\frac{\sqrt{w_{\phi}}}{|\sin\alpha\sin\theta|})^2.
\endaligned
\end{eqnarray}
Thus
\begin{eqnarray}
\aligned
&\left( \frac{w_{\alpha}}{\left(N+\sum\limits_{j\neq k}r^{(j,k)}_{\alpha\alpha}\right)}+\frac{w_{\theta}/\sin^2\alpha}{\left(N+\sum\limits_{j\neq k}r^{(j,k)}_{\theta\theta}\right)}+\frac{w_{\phi}/\sin^2\alpha\sin^2\theta}{\left(N+\sum\limits_{j\neq k}r^{(j,k)}_{\phi\phi}\right)}\right)\\
\geq & \frac{(\sqrt{w_{\alpha}}+\frac{\sqrt{w_{\theta}}}{|\sin\alpha|}|+\frac{\sqrt{w_{\phi}}}{|\sin\alpha\sin\theta|})^2}{\left(N+\sum\limits_{j\neq k}r^{(j,k)}_{\alpha\alpha}+N+\sum\limits_{j\neq k}r^{(j,k)}_{\theta\theta}+N+\sum\limits_{j\neq k}r^{(j,k)}_{\phi\phi}\right)}\\
\geq &\frac{(\sqrt{w_{\alpha}}+\frac{\sqrt{w_{\theta}}}{|\sin\alpha|}+\frac{\sqrt{w_{\phi}}}{|\sin\alpha\sin\theta|})^2}{N(N+2)},
\endaligned
\end{eqnarray}
where for the second inequality we used the fact that $r^{(j,k)}_{\alpha\alpha}+r^{(j,k)}_{\theta\theta}+r^{(j,k)}_{\phi\phi}\leq 1$ for $\forall j\neq k$, thus $\sum\limits_{j\neq k}r^{(j,k)}_{\alpha\alpha}+r^{(j,k)}_{\theta\theta}+r^{(j,k)}_{\phi\phi}\leq N(N-1)$.
The lower bound on the figure of merit can then be obtained as
\begin{equation}\label{eq:optimalbound}
w_{\alpha}\delta \hat{\alpha}^2+w_{\theta}\delta \hat{\theta}^2+w_{\phi}\delta \hat{\phi}^2\geq \frac{(\sqrt{w_{\alpha}}+\frac{\sqrt{w_{\theta}}}{|\sin\alpha|}|+\frac{\sqrt{w_{\phi}}}{|\sin\alpha\sin\theta|})^2}{4N(N+2)},
\end{equation}
which can be saturated when
\begin{eqnarray}\label{eq:optimalcoeff}
\aligned
&r_{\alpha\alpha}^{(j,k)}=\tilde{r}_{\alpha\alpha}=\frac{(N+1)\sqrt{w_{\alpha}}-\frac{\sqrt{w_{\theta}}}{|\sin\alpha|}-\frac{\sqrt{w_{\phi}}}{|\sin\alpha\sin\theta|}}{(N-1)(\sqrt{w_{\alpha}}+\frac{\sqrt{w_{\theta}}}{|\sin\alpha|}+\frac{\sqrt{w_{\phi}}}{|\sin\alpha\sin\theta|})},\\ &r_{\theta\theta}^{(j,k)}=\tilde{r}_{\theta\theta}=\frac{(N+1)\frac{\sqrt{w_{\theta}}}{|\sin\alpha|}-\sqrt{w_{\alpha}}-\frac{\sqrt{w_{\phi}}}{|\sin\alpha\sin\theta|}}{(N-1)(\sqrt{w_{\alpha}}+\frac{\sqrt{w_{\theta}}}{|\sin\alpha|}+\frac{\sqrt{w_{\phi}}}{|\sin\alpha\sin\theta|})},\\ &r_{\phi\phi}^{(j,k)}=\tilde{r}_{\phi\phi}=\frac{(N+1)\frac{\sqrt{w_{\phi}}}{|\sin\alpha\sin\theta|}-\sqrt{w_{\alpha}}-\frac{\sqrt{w_{\theta}}}{|\sin\alpha|}}{(N-1)(\sqrt{w_{\alpha}}+\frac{\sqrt{w_{\theta}}}{|\sin\alpha|}+\frac{\sqrt{w_{\phi}}}{|\sin\alpha\sin\theta|})},
\endaligned
\end{eqnarray}
and $\sum\limits_{k=0}^{N-1}r_\alpha^{(k)}=\sum\limits_{k=0}^{N-1}r_\phi^{(k)}=\sum\limits_{k=0}^{N-1}r_\phi^{(k)}=0$.

\section{Optimal probe state with the ancillary system \label{Sec:app.ancilla}}
With a three-level ancillary system, we can prepare the probe state as
\begin{equation}
    | \Psi_{SA} \rangle=s_\alpha | \Phi_{\alpha} \rangle\otimes\ket{0}+s_\theta | \Phi_{\theta} \rangle\otimes\ket{1}+s_\phi| \Phi_{\phi} \rangle\otimes\ket{2},
\end{equation}
with $| \Phi_{x} \rangle=\frac{1}{\sqrt{2}}\left( |+_{x}\right\rangle^{\otimes N}+|-_{x} \rangle^{\otimes N} )$ for $x\in\{\alpha,\theta,\phi\}$, and $\ket{\pm_x}$ are the eigenvectors of $\bm{n}_x\cdot\bm{\sigma}$, $\{\ket{0},\ket{1},\ket{2}\}$ is an orthonormal basis of the ancillary system. The normalization condition requires that $|s_\alpha|^2+|s_\theta|^2+|s_\phi|^2=1$.
The entries of the quantum Fisher information matrix can be obtained as
\begin{align}
    J_{x, y \in\{\alpha, \theta, \phi\}} =&2\left\langle\Psi_{SA}\left|H_{x}^{(N)} H_{y}^{(N)}+H_{y}^{(N)}H_{x}^{(N)}\right| \Psi_{SA}\right\rangle \\
    &- 4\left\langle\Psi_{SA}\left|H_{x}^{(N)}\right| \Psi_{SA}\right\rangle\left\langle\Psi_{SA}\left|H_{y}^{(N)}\right| \Psi_{SA}\right\rangle \\
    =&4c_xc_y\left[N\delta _{xy} + \frac{N(N-1)}{2}(r_{x y}+r_{y x})-N^2(r_{x 0}r_{y 0})\right],
\end{align}
here
\begin{equation}
    r_{xy}=\bra{\Psi_{SA}}\bm{n}_x\cdot \bm{\sigma} \otimes \bm{n}_y\cdot \bm{\sigma}\ket{\Psi_{SA}}=\tr\left[\rho^{[2]}(\bm{n}_x\cdot\bm{\sigma}\otimes\bm{n}_y\cdot\bm{\sigma})\right],
\end{equation}
\begin{equation}
    r_{x0}=\bra{\Psi_{SA}}\bm{n}_x\cdot \bm{\sigma}\ket{\Psi_{SA}}=\tr\left[\rho^{[1]}(\bm{n}_x\cdot\bm{\sigma})\right],
\end{equation}
where $\rho^{[2]}$ is the reduced two-spin state of $| \Psi_{SA} \rangle=s_\alpha | \Phi_{\alpha} \rangle\otimes\ket{0}+s_\theta | \Phi_{\theta} \rangle\otimes\ket{1}+s_\phi| \Phi_{\phi} \rangle\otimes\ket{2}$, which is
\begin{eqnarray}\label{eq:reducedtwo}
\rho^{[2]}=\frac{1}{4}[I+r_{\alpha \alpha}\sigma_\alpha\otimes\sigma_\alpha+r_{\theta \theta}\sigma_\theta\otimes\sigma_\theta+r_{\phi \phi}\sigma_\phi\otimes\sigma_\phi]
\end{eqnarray}
with $r_{\alpha \alpha}=|s_\alpha|^2$, $r_{\theta \theta}=|s_\theta|^2$,$r_{\phi \phi}=|s_\phi|^2$,
$\rho^{[1]}$ is the reduced single spin state of $|\Psi_{SA}\rangle$, which is $\rho^{[1]}=I/2$.

Thus in this case $r_{\alpha 0}=r_{\theta 0}=r_{\phi 0}=0$, $r_{\alpha\theta}=r_{\alpha\phi}=r_{\theta\phi}=0$, $r_{\alpha\alpha}=|s_\alpha|^2$, $r_{\theta\theta}=|s_\theta|^2$, $r_{\phi\phi}=|s_\phi|^2$, from which we can obtain the quantum Fisher information matrix as
\begin{equation}\label{eq:QFIMapp}
    J=4N J_1+4N(N-1) J_2,
\end{equation}
with
\begin{equation}
    J_{1} \quad=\left( \begin{array}{ccc}{1} & {0} & {0} \\ {0} & {\sin ^{2} \alpha} & {0} \\ {0} & {0} & {\sin ^{2} \alpha \sin ^{2} \theta}\end{array}\right),
\end{equation}
\begin{equation}
    J_{2}=\left( \begin{array}{ccc}{|s_\alpha|^2} & {0} & {0} \\ {0} & {|s_\theta|^2 \sin ^{2} \alpha} & {0} \\ {0} & {0} & {|s_\phi|^2 \sin ^{2} \alpha \sin ^{2} \theta}\end{array}\right).
\end{equation}

When $\tilde{r}_{\alpha\alpha}$, $\tilde{r}_{\theta\theta}$, and $\tilde{r}_{\phi\phi}$ in Eq.(\ref{eq:optimalcoeff}) are all non-negative(which always hold for sufficient large $N$), we can take
$s_\alpha=\sqrt{\tilde{r}_{\alpha\alpha}}$, $s_\theta=\sqrt{\tilde{r}_{\theta\theta}}$, $s_\phi=\sqrt{\tilde{r}_{\phi\phi}}$. The ultimate lower bound in Eq.(\ref{eq:optimalbound}) is saturated.

If $\tilde{r}_{\alpha\alpha}$, $\tilde{r}_{\theta\theta}$, and $\tilde{r}_{\phi\phi}$ are not all non-negative, then we need to optimize the coefficients of $|\Phi_{SA} \rangle$ to find the best precision achievable by this state.
%
From the QFIM given in Eq.(\ref{eq:QFIMapp}), we can obtain the QCRB for the weighted sum of variances as
\begin{equation}\label{eq.QCRB}
  \begin{aligned}
    &w_{\alpha} \delta \hat{\alpha}^{2}+w_{\theta} \delta \hat{\theta}^{2}+w_{\phi} \delta \hat{\phi}^{2}\ge \frac{1}{4N}\left( \frac{w_{\alpha}}{1+(N-1)r_{\alpha\alpha}}+\frac{w_{\theta} / \sin ^{2} \alpha}{1+(N-1)r_{\theta\theta}}+\frac{w_{\phi} / \sin ^{2} \alpha \sin ^{2} \theta}{1+(N-1)r_{\phi\phi}} \right),
  \end{aligned}
\end{equation}
which can be saturated as the weak commutativity condition holds. To find the best precision, we just need to find the optimal coefficients such that the right side of the above equation, which we denote as $f=\frac{1}{4N}\left( \frac{w_{\alpha}}{1+(N-1)r_{\alpha\alpha}}+\frac{w_{\theta} / \sin ^{2} \alpha}{1+(N-1)r_{\theta\theta}}+\frac{w_{\phi} / \sin ^{2} \alpha \sin ^{2} \theta}{1+(N-1)r_{\phi\phi}} \right)$, is minimized. Since $r_{\alpha\alpha}+r_{\theta\theta}+r_{\phi\phi}=1$, we can view $f$ as a two-variable function $f(r_{\alpha\alpha},r_{\theta\theta})$ by replacing $r_{\phi\phi}$ with $1-r_{\alpha\alpha}-r_{\theta\theta}$.
The ultimate precision attainable with $|\Phi_{SA}\rangle$ is then the minimum of $f(r_{\alpha\alpha},r_{\theta\theta})$ under the contraints $r_{\alpha\alpha},r_{\theta\theta}\ge 0$ and $r_{\alpha\alpha}+r_{\theta\theta}\le 1$.
We first note that $f(r_{\alpha\alpha},r_{\theta\theta})$ is convex in its domain since the Hessian $\mathcal{H}$ is positive definite which can be seen as
\begin{equation}
  \mathcal{H}_{11}=\frac{\partial^2f}{\partial r_{\alpha\alpha}^2}=\frac{(N-1)^2}{2N} \frac{w_{\alpha}}{[1+(N-1) r_{\alpha \alpha}]^3}>0,
\end{equation}
\begin{equation}
  \mathcal{H}_{22}=\frac{\partial^2f}{\partial r_{\theta\theta}^2}=\frac{(N-1)^2}{2N} \frac{w_{\theta} / \sin ^{2} \alpha}{[1+(N-1) r_{\theta \theta}]^3}>0,
\end{equation}
\begin{equation}
  \begin{aligned}
    \mathrm{det}\mathcal{H}=&\frac{\partial^2f}{\partial r_{\alpha\alpha}^2}\frac{\partial^2f}{\partial r_{\theta\theta}^2}-\left( \frac{\partial^2 f}{\partial r_{\alpha\alpha}\partial r_{\theta\theta}} \right)^2\\
    =&\frac{(N-1)^4}{4N^2}\Bigg( \frac{w_{\alpha}}{[1+(N-1) r_{\alpha \alpha}]^3}\frac{w_{\theta} / \sin ^{2} \alpha}{[1+(N-1) r_{\theta \theta}]^3}\Bigg.\\
    &\Bigg.+\left(  \frac{w_{\alpha}}{[1+(N-1) r_{\alpha \alpha}]^3}+\frac{w_{\theta} / \sin ^{2} \alpha}{[1+(N-1) r_{\theta \theta}]^3} \right) \frac{w_{\phi} / \sin ^{2} \alpha \sin ^{2} \theta}{[1+(N-1) r_{\phi \phi}]^3} \Bigg)>0.
  \end{aligned}
\end{equation}
The local minimum of $f(r_{\alpha\alpha},r_{\theta\theta})$ is thus also the global minimum.

We note that $\partial f/\partial r_{\alpha\alpha}=\partial f/\partial r_{\theta\theta}=0$ has only one solution, which is given by the values in Eq.(\ref{eq:optimalcoeff}). 
If some of the values in Eq.(\ref{eq:optimalcoeff}) are negative, i.e., at least one of the conditions
\begin{equation}
  \begin{aligned}
    \mathrm{I(a)}:& (N+1) \sqrt{w_{\alpha}}\ge \frac{\sqrt{w_{\theta}}}{|\sin \alpha|}+\frac{\sqrt{w_{\phi}}}{|\sin \alpha \sin \theta|}\\
    \mathrm{I(b)}:& (N+1) \frac{\sqrt{w_{\theta}}}{|\sin \alpha|}\ge \sqrt{w_{\alpha}}+\frac{\sqrt{w_{\phi}}}{|\sin \alpha \sin \theta|}\\
    \mathrm{I(c)}:& (N+1) \frac{\sqrt{w_{\phi}}}{|\sin \alpha \sin \theta|}\ge \sqrt{w_{\alpha}}+\frac{\sqrt{w_{\theta}}}{|\sin \alpha|}
  \end{aligned}
\end{equation}
fails, then we need to consider the points at the boundary of the domain since the extreme point is out of the feasible domain.
For example, at the boundary of $r_{\alpha\alpha}=0$, we need to compare two end points
\begin{equation}
  f(0,0)=\frac{1}{4N}\left( w_{\alpha}+w_{\theta} / \sin ^{2} \alpha+\frac{w_{\phi} / \sin ^{2} \alpha \sin ^{2} \theta}{N} \right)
\end{equation}
\begin{equation}
  f(0,1)=\frac{1}{4N}\left( w_{\alpha}+\frac{w_{\theta} / \sin ^{2} \alpha}{N}+w_{\phi} / \sin ^{2} \alpha \sin ^{2} \theta \right)
\end{equation}
and one extreme point on this boundary given by $\partial f/\partial r_{\theta\theta}=0$, which is given by $r_{\theta\theta}^*=\frac{N\frac{\sqrt{w_{\theta}}}{|\sin \alpha|}-\frac{\sqrt{w_{\phi}}}{|\sin \alpha \sin \theta|}}{(N-1)\left(\frac{\sqrt{w_{\theta}}}{|\sin \alpha|}+\frac{\sqrt{w_{\phi}}}{|\sin \alpha \sin \theta|}\right)}$ and
\begin{equation}
  f(0,r_{\theta\theta}^*)=\frac{1}{4N}\left( w_{\alpha}+\frac{\left(\frac{\sqrt{w_{\theta}}}{|\sin \alpha|}+\frac{\sqrt{w_{\phi}}}{|\sin \alpha \sin \theta|}\right)^2}{N+1} \right).
\end{equation}
We note that $(0,r_{\theta\theta}^*)$ is in the feasible domain if and only if $\frac{1}{N}\frac{\sqrt{w_{\theta}}}{|\sin \alpha|}\le \frac{\sqrt{w_{\phi}}}{|\sin \alpha \sin \theta|}\le N\frac{\sqrt{w_{\theta}}}{|\sin \alpha|}$ and when it is in the feasible domain $f(0,r_{\theta\theta}^*)$ is smaller than $\min\{f(0,0),f(0,1)\}$.
Similarly, one can find the special points at the boundary of $r_{\theta\theta}=0$ as
\begin{equation}
  f(1,0)=\frac{1}{4N}\left( \frac{w_{\alpha}}{N}+w_{\theta} / \sin ^{2} \alpha+w_{\phi} / \sin ^{2} \alpha \sin ^{2} \theta \right),
\end{equation}
\begin{equation}
  f(r_{\alpha\alpha}^*,0)=f(\frac{N\sqrt{w_{\alpha}}-\frac{\sqrt{w_{\phi}}}{|\sin \alpha \sin \theta|}}{(N-1)\left(\sqrt{w_{\alpha}}+\frac{\sqrt{w_{\phi}}}{|\sin \alpha \sin \theta|}\right)},0)=\frac{1}{4N}\left( \frac{w_{\theta}}{\sin^2 \alpha}+\frac{\left(\sqrt{w_{\alpha}}+\frac{\sqrt{w_{\phi}}}{|\sin \alpha \sin \theta|}\right)^2}{N+1} \right),
\end{equation}
where $(r_{\alpha\alpha}^*,0)$ is in the domain if and only if $\frac{1}{N}\sqrt{w_\alpha}\le\frac{\sqrt{w_{\phi}}}{|\sin \alpha \sin \theta|}\le N\sqrt{w_{\alpha}}$). The special points at the boundary of $r_{\alpha\alpha}+r_{\theta\theta}=1$ are
\begin{equation}
  \begin{aligned}
     f(r_{\alpha\alpha}',r_{\theta\theta}')=&f(\frac{N\sqrt{w_{\alpha}}-\frac{\sqrt{w_{\theta}}}{|\sin \alpha|}}{(N-1)\left(\sqrt{w_{\alpha}}+\frac{\sqrt{w_{\theta}}}{|\sin \alpha|}\right)},\frac{-\sqrt{w_{\alpha}}+N\frac{\sqrt{w_{\theta}}}{|\sin \alpha|}}{(N-1)\left(\sqrt{w_{\alpha}}+\frac{\sqrt{w_{\theta}}}{|\sin \alpha|}\right)})
     \\=&\frac{1}{4N}\left( \frac{w_{\phi}}{\sin^2 \alpha \sin^2 \theta} +\frac{\left(\frac{\sqrt{w_{\theta}}}{|\sin \alpha|}+\sqrt{w_{\alpha}}\right)^2}{N+1} \right)
  \end{aligned}
\end{equation}
where $(r_{\alpha\alpha}',r_{\theta\theta}')$ is in the feasible domain if and only if $\frac{1}{N}\sqrt{w_\alpha}\le \frac{\sqrt{w_{\theta}}}{|\sin \alpha|}\le N\sqrt{w_\alpha}$.

Let
\begin{equation}
  \begin{aligned}
    \mathrm{II(a)}:&\frac{1}{N}\frac{\sqrt{w_{\theta}}}{|\sin \alpha|}\le \frac{\sqrt{w_{\phi}}}{|\sin \alpha \sin \theta|}\le N\frac{\sqrt{w_{\theta}}}{|\sin \alpha|}\\
    \mathrm{II(b)}:&\frac{1}{N}\sqrt{w_\alpha}\le \frac{\sqrt{w_{\phi}}}{|\sin \alpha \sin \theta|}\le N\sqrt{w_\alpha}\\
    \mathrm{II(c)}:&\frac{1}{N}\sqrt{w_\alpha}\le \frac{\sqrt{w_{\theta}}}{|\sin \alpha|}\le N\sqrt{w_\alpha}.
  \end{aligned}
\end{equation}
the minimal sum of weighted variance achievable by $|\Psi_{SA}\rangle$ can be obtained as following:
\begin{itemize}
  \item If I(a)-(c) hold, $f_{\min}=\frac{\left(\sqrt{w_{\alpha}}+\frac{\sqrt{w_{\theta}}}{|\sin \alpha|}+\frac{\sqrt{w_{\phi}}}{|\sin \alpha \sin \theta|}\right)^{2}}{4 N(N+2)}$.
  \item If at least one of I(a)-(c) is false:
   \begin{itemize}
   \item II(a)-(c) all hold,
  then $f_{\min}=\min\{f(0,r_{\theta\theta}^*),f(r_{\alpha\alpha}^*,0),f(r_{\alpha\alpha}',r_{\theta\theta}')\}$, here $r_{\theta\theta}^*=\frac{N\frac{\sqrt{w_{\theta}}}{|\sin \alpha|}-\frac{\sqrt{w_{\phi}}}{|\sin \alpha \sin \theta|}}{(N-1)\left(\frac{\sqrt{w_{\theta}}}{|\sin \alpha|}+\frac{\sqrt{w_{\phi}}}{|\sin \alpha \sin \theta|}\right)}$, $r_{\alpha\alpha}^*=\frac{N\sqrt{w_{\alpha}}-\frac{\sqrt{w_{\phi}}}{|\sin \alpha \sin \theta|}}{(N-1)\left(\sqrt{w_{\alpha}}+\frac{\sqrt{w_{\phi}}}{|\sin \alpha \sin \theta|}\right)}$,
  $r_{\alpha\alpha}'=\frac{N\sqrt{w_{\alpha}}-\frac{\sqrt{w_{\theta}}}{|\sin \alpha|}}{(N-1)\left(\sqrt{w_{\alpha}}+\frac{\sqrt{w_{\theta}}}{|\sin \alpha|}\right)}$,$r_{\theta\theta}'=\frac{-\sqrt{w_{\alpha}}+N\frac{\sqrt{w_{\theta}}}{|\sin \alpha|}}{(N-1)\left(\sqrt{w_{\alpha}}+\frac{\sqrt{w_{\theta}}}{|\sin \alpha|}\right)}$.
  \item only II(a) is false: $f_{\min}=\min\{f(r_{\alpha\alpha}^*,0),f(r_{\alpha\alpha}',r_{\theta\theta}')\}$.
  \item only II(b) is false: $f_{\min}=\min\{f(0,r_{\theta\theta}^*),f(r_{\alpha\alpha}',r_{\theta\theta}')\}$.
  \item only II(c) is false:
  $f_{\min}=\min\{f(0,r_{\theta\theta}^*),f(r_{\alpha\alpha}^*,0))\}$.
  \item only II(a) holds:
  $f_{\min}=\min\{f(0,r_{\theta\theta}^*),f(1,0)\}$

  \item only II(b) holds:
  $f_{\min}=\min\{ f(r_{\alpha\alpha}^*,0), f(0,1)\}$

  \item only II(c) holds:
  $f_{\min}=\min\{ f(r_{\alpha\alpha}',r_{\theta\theta}'), f(0,0)\}$

  \item II(a-c) are all false, then $f_{\min}=\min\{f(0,0),f(0,1),f(1,0)\}$.
   \end{itemize}
\end{itemize}

\section{Probe state without ancillary system \label{app.B}}
For finite N, we consider the state
\begin{eqnarray}
\label{eq:probeNapp}
\aligned
     | \Phi_{o} \rangle=\frac{1}{\sqrt{2}M}[&s_{\alpha}(\ket{+_\alpha}^{\otimes N}+e^{i\gamma_1}\ket{-_\alpha}^{\otimes N})\\
&+s_{\theta}(e^{i\gamma_2}\ket{+_\theta}^{\otimes N}+e^{i\gamma_3}\ket{-_\theta}^{\otimes N})\\
&+s_{\phi}(e^{i\gamma_4}\ket{+_\phi}^{\otimes N}+e^{i\gamma_5}\ket{-_\phi}^{\otimes N})].
\endaligned
\end{eqnarray}
We can write
\begin{align}
  \sigma_1 &=\ket{+_x}\bra{+_x}-\ket{-_x}\bra{-_x}, \\
  \sigma_2 &=\ket{+_y}\bra{+_y}-\ket{-_y}\bra{-_y}, \\
  \sigma_3 &=\ket{+_z}\bra{+_z}-\ket{-_z}\bra{-_z},
\end{align}
with
\begin{align*}
    \ket{+_x}=\frac{1}{\sqrt{2}}\begin{bmatrix}
        1 \\ 1
\end{bmatrix}, &\hspace{2cm} \ket{+_y}=\frac{1}{\sqrt{2}}\begin{bmatrix}
    -i \\ 1
\end{bmatrix}, & \ket{+_z}=\begin{bmatrix}
    1 \\ 0
\end{bmatrix}, \\
\ket{-_x}=\frac{1}{\sqrt{2}}\begin{bmatrix}
    -1 \\ 1
\end{bmatrix}, &\hspace{2cm} \ket{-_y}=\frac{1}{\sqrt{2}}\begin{bmatrix}
i \\ 1
\end{bmatrix}, & \ket{-_z}=\begin{bmatrix}
0 \\ 1
\end{bmatrix}.
\end{align*}
There exists a unitary transformation $U=e^{-i\frac{\alpha}{2}\bm{n}\cdot\bm{\sigma}}e^{-i\frac{\phi}{2}\sigma_3}e^{-i\frac{\theta}{2}\sigma_2}$ such that
\begin{align}
    \bm{n}_{\theta}\cdot\bm{\sigma} &= U\sigma_1 U^{\dagger}, \\
    \bm{n}_{\phi}\cdot\bm{\sigma} &= U\sigma_2 U^{\dagger}, \\
    \bm{n}_{\alpha}\cdot\bm{\sigma} &= U\sigma_3 U^{\dagger}.
\end{align}
Therefore, it is easy to obtain
\begin{align*}
    \ket{+_\theta}=U\frac{1}{\sqrt{2}}\begin{bmatrix}
        1 \\ 1
\end{bmatrix}, &\hspace{2cm} \ket{+_\phi}=U\frac{1}{\sqrt{2}}\begin{bmatrix}
    -i \\ 1
\end{bmatrix}, & \ket{+_\alpha}=U\begin{bmatrix}
    1 \\ 0
\end{bmatrix}, \\
\ket{-_\theta}=U\frac{1}{\sqrt{2}}\begin{bmatrix}
    -1 \\ 1
\end{bmatrix}, &\hspace{2cm} \ket{-_\phi}=U\frac{1}{\sqrt{2}}\begin{bmatrix}
i \\ 1
\end{bmatrix}, & \ket{-_\alpha}=U\begin{bmatrix}
0 \\ 1
\end{bmatrix}.
\end{align*}
By taking the phases into consideration, the normalization constant equals to
\begin{align}
    M^2 &=s^2_{\alpha}+s^2_{\theta}+s^2_{\phi} \\\nonumber
    {} &  + s_{\alpha}s_{\theta} \left[ (\frac{1}{\sqrt{2}})^N(\frac{e^{i\gamma_2}+e^{-i\gamma_2}}{2}+\frac{e^{i(\gamma_2-\gamma_1)}+e^{-i(\gamma_2-\gamma_1)}}{2}+\frac{e^{i(\gamma_3-\gamma_1)}+e^{-i(\gamma_3-\gamma_1)}}{2})\right.\\\nonumber
    {}&\hspace{5cm}+\left.(\frac{-1}{\sqrt{2}})^N(\frac{e^{i\gamma_3}+e^{-i\gamma_3}}{2}) \right] \\\nonumber
    {} & + s_{\alpha}s_{\phi}\left[ (\frac{1}{\sqrt{2}})^N(\frac{e^{i(\gamma_4-\gamma_1)}+e^{-i(\gamma_4-\gamma_1)}}{2}+\frac{e^{i(\gamma_5-\gamma_1)}+e^{-i(\gamma_5-\gamma_1)}}{2})\right.\\\nonumber
    {} &\hspace{5cm} +\left.(\frac{i}{\sqrt{2}})^N(\frac{e^{i\gamma_5}+e^{-i\gamma_4}}{2})+(\frac{-i}{\sqrt{2}})^N(\frac{e^{i\gamma_4}+e^{-i\gamma_5}}{2})\right] \\\nonumber
    {} &  + s_{\theta}s_{\phi}\left[ (\frac{1+i}{2})^N(\frac{e^{i(\gamma_4-\gamma_3)}+e^{i(\gamma_5-\gamma_2)}+e^{-i(\gamma_4-\gamma_2)}+e^{-i(\gamma_5-\gamma_3)}}{2})\right.\\\nonumber
    {} &\hspace{5cm}+\left.(\frac{1-i}{2})^N(\frac{e^{i(\gamma_4-\gamma_2)}+e^{i(\gamma_5-\gamma_3)}+e^{-i(\gamma_4-\gamma_3)}+e^{-i(\gamma_5-\gamma_2)}}{2}) \right].
\end{align}
We write the reduced two-qubit state as $\rho^{[2]}=\frac{1}{4}[I+\sum_{x}r_{x0}(\sigma_x\otimes I+I\otimes \sigma_x)+\sum_{x,y}r_{xy}\sigma_x\otimes\sigma_y]$, here $x,y\in \{\alpha,\theta,\phi\}$ and $r_{xy}=r_{yx}$, with $r_{xy}=\bra{\Phi_o}\bm{n}_x\cdot\bm{\sigma}\otimes\bm{n}_y\cdot\bm{\sigma}\ket{\Phi_o}$ given as
\begin{align}
    r_{\alpha\alpha} &=\bra{\Phi_o}\bm{n}_\alpha\cdot\bm{\sigma}\otimes\bm{n}_\alpha\cdot\bm{\sigma}\ket{\Phi_o} \\\nonumber &=\frac{1}{M^2}\left\{s^2_{\alpha}+s_{\alpha}s_{\theta} \left[ (\frac{1}{\sqrt{2}})^N(\frac{e^{i\gamma_2}+e^{-i\gamma_2}}{2}+\frac{e^{i(\gamma_2-\gamma_1)}+e^{-i(\gamma_2-\gamma_1)}}{2}+\frac{e^{i(\gamma_3-\gamma_1)}+e^{-i(\gamma_3-\gamma_1)}}{2})\right.\right.\\\nonumber
    {}&\hspace{5cm}+\left.(\frac{-1}{\sqrt{2}})^N(\frac{e^{i\gamma_3}+e^{-i\gamma_3}}{2}) \right] \\\nonumber
    {} & + s_{\alpha}s_{\phi}\left[ (\frac{1}{\sqrt{2}})^N(\frac{e^{i(\gamma_4-\gamma_1)}+e^{-i(\gamma_4-\gamma_1)}}{2}+\frac{e^{i(\gamma_5-\gamma_1)}+e^{-i(\gamma_5-\gamma_1)}}{2})\right.\\\nonumber
    {} &\hspace{5cm} +\left.(\frac{i}{\sqrt{2}})^N(\frac{e^{i\gamma_5}+e^{-i\gamma_4}}{2})+(\frac{-i}{\sqrt{2}})^N(\frac{e^{i\gamma_4}+e^{-i\gamma_5}}{2})\right] \\\nonumber
    {} &  + s_{\theta}s_{\phi}\left[ (\frac{1+i}{2})^{N-2}(\frac{1-i}{2})^2(\frac{e^{i(\gamma_4-\gamma_3)}+e^{-i(\gamma_4-\gamma_2)}+e^{i(\gamma_5-\gamma_2)}+e^{-i(\gamma_5-\gamma_3)}}{2})\right.\\\nonumber
    {} &\hspace{5cm}+\left.\left.(\frac{1-i}{2})^{N-2}(\frac{1+i}{2})^2(\frac{e^{i(\gamma_4-\gamma_2)}+e^{-i(\gamma_4-\gamma_3)}+e^{i(\gamma_5-\gamma_3)}+e^{-i(\gamma_5-\gamma_2)}}{2}) \right]\right\},
\end{align}

\begin{align}
    r_{\theta\theta} &=\bra{\Phi_o}\bm{n}_\theta\cdot\bm{\sigma}\otimes\bm{n}_\theta\cdot\bm{\sigma}\ket{\Phi_o} \\\nonumber &=\frac{1}{M^2}\left\{s^2_{\theta}+s_{\alpha}s_{\theta} \left[ (\frac{1}{\sqrt{2}})^N(\frac{e^{i\gamma_2}+e^{-i\gamma_2}}{2}+\frac{e^{i(\gamma_2-\gamma_1)}+e^{-i(\gamma_2-\gamma_1)}}{2}+\frac{e^{i(\gamma_3-\gamma_1)}+e^{-i(\gamma_3-\gamma_1)}}{2})\right.\right.\\\nonumber
    {}&\hspace{5cm}+\left.(\frac{-1}{\sqrt{2}})^N(\frac{e^{i\gamma_3}+e^{-i\gamma_3}}{2}) \right] \\\nonumber
    {} & + s_{\alpha}s_{\phi}\left[ (\frac{1}{\sqrt{2}})^{N-2}(-\frac{e^{i(\gamma_4-\gamma_1)}+e^{-i(\gamma_4-\gamma_1)}+e^{i(\gamma_5-\gamma_1)}+e^{-i(\gamma_5-\gamma_1)}}{4})\right.\\\nonumber
    {} &\hspace{5cm} +\left.(\frac{i}{\sqrt{2}})^{N-2}(\frac{e^{i\gamma_5}+e^{-i\gamma_4}}{4})+(\frac{-i}{\sqrt{2}})^{N-2}(\frac{e^{i\gamma_4}+e^{-i\gamma_5}}{4})\right] \\\nonumber
    {} &  + s_{\theta}s_{\phi}\left[ (\frac{1+i}{2})^N(\frac{e^{i(\gamma_4-\gamma_3)}+e^{i(\gamma_5-\gamma_2)}+e^{-i(\gamma_4-\gamma_2)}+e^{-i(\gamma_5-\gamma_3)}}{2})\right.\\\nonumber
    {} &\hspace{5cm}+\left.\left.(\frac{1-i}{2})^N(\frac{e^{i(\gamma_4-\gamma_2)}+e^{i(\gamma_5-\gamma_3)}+e^{-i(\gamma_4-\gamma_3)}+e^{-i(\gamma_5-\gamma_2)}}{2}) \right]\right\},
\end{align}

\begin{align}
    r_{\phi\phi} &=\bra{\Phi_o}\bm{n}_\phi\cdot\bm{\sigma}\otimes\bm{n}_\phi\cdot\bm{\sigma}\ket{\Phi_o} \\\nonumber &=\frac{1}{M^2}\left\{s^2_{\phi}+s_{\alpha}s_{\theta} \left[ -(\frac{1}{\sqrt{2}})^N(\frac{e^{i\gamma_2}+e^{-i\gamma_2}}{2}+\frac{e^{i(\gamma_2-\gamma_1)}+e^{-i(\gamma_2-\gamma_1)}}{2}+\frac{e^{i(\gamma_3-\gamma_1)}+e^{-i(\gamma_3-\gamma_1)}}{2})\right.\right.\\\nonumber
    {}&\hspace{5cm}-\left.(\frac{-1}{\sqrt{2}})^N(\frac{e^{i\gamma_3}+e^{-i\gamma_3}}{2}) \right] \\\nonumber
    {} & + s_{\alpha}s_{\phi}\left[ (\frac{1}{\sqrt{2}})^N(\frac{e^{i(\gamma_4-\gamma_1)}+e^{-i(\gamma_4-\gamma_1)}}{2}+\frac{e^{i(\gamma_5-\gamma_1)}+e^{-i(\gamma_5-\gamma_1)}}{2})\right.\\\nonumber
    {} &\hspace{5cm} +\left.(\frac{i}{\sqrt{2}})^N(\frac{e^{i\gamma_5}+e^{-i\gamma_4}}{2})+(\frac{-i}{\sqrt{2}})^N(\frac{e^{i\gamma_4}+e^{-i\gamma_5}}{2})\right] \\\nonumber
    {} &  + s_{\theta}s_{\phi}\left[ (\frac{1+i}{2})^N(\frac{e^{i(\gamma_4-\gamma_3)}+e^{i(\gamma_5-\gamma_2)}+e^{-i(\gamma_4-\gamma_2)}+e^{-i(\gamma_5-\gamma_3)}}{2})\right.\\\nonumber
    {} &\hspace{5cm}+\left.\left.(\frac{1-i}{2})^N(\frac{e^{i(\gamma_4-\gamma_2)}+e^{i(\gamma_5-\gamma_3)}+e^{-i(\gamma_4-\gamma_3)}+e^{-i(\gamma_5-\gamma_2)}}{2}) \right]\right\},
\end{align}

\begin{align}
    r_{\alpha\theta} &=\bra{\Phi_o}\bm{n}_\alpha\cdot\bm{\sigma}\otimes\bm{n}_\theta\cdot\bm{\sigma}\ket{\Phi_o} \\\nonumber &=\frac{1}{M^2}\left\{s_{\alpha}s_{\theta} \left[ (\frac{1}{\sqrt{2}})^N(\frac{e^{i\gamma_2}+e^{-i\gamma_2}}{2}-\frac{e^{i(\gamma_2-\gamma_1)}+e^{-i(\gamma_2-\gamma_1)}}{2}+\frac{e^{i(\gamma_3-\gamma_1)}+e^{-i(\gamma_3-\gamma_1)}}{2})\right.\right.\\\nonumber
    {}&\hspace{5cm}-\left.(\frac{-1}{\sqrt{2}})^N(\frac{e^{i\gamma_3}+e^{-i\gamma_3}}{2}) \right] \\\nonumber
    {} & + s_{\alpha}s_{\phi}\left[ (\frac{1}{\sqrt{2}})^{N-1}\frac{i}{\sqrt{2}}(\frac{e^{i(\gamma_4-\gamma_1)}-e^{-i(\gamma_4-\gamma_1)}-e^{i(\gamma_5-\gamma_1)}+e^{-i(\gamma_5-\gamma_1)}}{2})\right.\\\nonumber
    {} &\hspace{5cm} +\left.(\frac{i}{\sqrt{2}})^{N-1}(\frac{e^{i\gamma_5}+e^{-i\gamma_4}}{2\sqrt{2}})+(\frac{-i}{\sqrt{2}})^{N-1}(\frac{e^{i\gamma_4}+e^{-i\gamma_5}}{2\sqrt{2}})\right] \\\nonumber
    {} &  + s_{\theta}s_{\phi}\left[ (\frac{1+i}{2})^{N-2}\frac{e^{i(\gamma_4-\gamma_3)}-e^{-i(\gamma_4-\gamma_2)}-e^{i(\gamma_5-\gamma_2)}+e^{-i(\gamma_5-\gamma_3)}}{4}\right.\\\nonumber
    {} &\hspace{5cm}+\left.\left.(\frac{1-i}{2})^{N-2}\frac{-e^{i(\gamma_4-\gamma_2)}+e^{-i(\gamma_4-\gamma_3)}+e^{i(\gamma_5-\gamma_3)}-e^{-i(\gamma_5-\gamma_2)}}{4} \right]\right\},
\end{align}

\begin{align}
    r_{\alpha\phi} &=\bra{\Phi_o}\bm{n}_\alpha\cdot\bm{\sigma}\otimes\bm{n}_\phi\cdot\bm{\sigma}\ket{\Phi_o} \\\nonumber &=\frac{1}{M^2}\left\{s_{\alpha}s_{\theta} \left[ (\frac{1}{\sqrt{2}})^{N-1}\frac{-i}{\sqrt{2}}(\frac{e^{i\gamma_2}-e^{-i\gamma_2}}{2}+\frac{e^{i(\gamma_2-\gamma_1)}-e^{-i(\gamma_2-\gamma_1)}}{2}+\frac{-e^{i(\gamma_3-\gamma_1)}+e^{-i(\gamma_3-\gamma_1)}}{2})\right.\right.\\\nonumber
    {}&\hspace{5cm}+\left.(\frac{-1}{\sqrt{2}})^{N-1}\frac{-i}{\sqrt{2}}(\frac{e^{i\gamma_3}-e^{-i\gamma_3}}{2}) \right] \\\nonumber
    {} & + s_{\alpha}s_{\phi}\left[ (\frac{1}{\sqrt{2}})^N(-\frac{e^{i(\gamma_4-\gamma_1)}+e^{-i(\gamma_4-\gamma_1)}}{2}+\frac{e^{i(\gamma_5-\gamma_1)}+e^{-i(\gamma_5-\gamma_1)}}{2})\right.\\\nonumber
    {} &\hspace{5cm} +\left.(\frac{i}{\sqrt{2}})^N(\frac{-e^{i\gamma_5}+e^{-i\gamma_4}}{2})+(\frac{-i}{\sqrt{2}})^N(\frac{e^{i\gamma_4}-e^{-i\gamma_5}}{2})\right] \\\nonumber
    {} &  + s_{\theta}s_{\phi}\left[ (\frac{1+i}{2})^{N-2}(\frac{-e^{i(\gamma_4-\gamma_3)}+e^{i(\gamma_5-\gamma_2)}-e^{-i(\gamma_4-\gamma_2)}+e^{-i(\gamma_5-\gamma_3)}}{4})\right.\\\nonumber
    {} &\hspace{5cm}+\left.\left.(\frac{1-i}{2})^{N-2}(\frac{-e^{i(\gamma_4-\gamma_2)}+e^{i(\gamma_5-\gamma_3)}-e^{-i(\gamma_4-\gamma_3)}+e^{-i(\gamma_5-\gamma_2)}}{4}) \right]\right\},
\end{align}

\begin{align}
    r_{\theta\phi} &=\bra{\Phi_o}\bm{n}_\theta\cdot\bm{\sigma}\otimes\bm{n}_\phi\cdot\bm{\sigma}\ket{\Phi_o} \\\nonumber &=\frac{1}{M^2}\left\{s_{\alpha}s_{\theta} \left[ (\frac{1}{\sqrt{2}})^{N-1}\frac{i}{\sqrt{2}}(\frac{-e^{i\gamma_2}+e^{-i\gamma_2}}{2}+\frac{e^{i(\gamma_2-\gamma_1)}-e^{-i(\gamma_2-\gamma_1)}}{2}+\frac{e^{i(\gamma_3-\gamma_1)}-e^{-i(\gamma_3-\gamma_1)}}{2})\right.\right.\\\nonumber
    {}&\hspace{5cm}+\left.(\frac{-1}{\sqrt{2}})^{N-1}\frac{i}{\sqrt{2}}(\frac{e^{i\gamma_3}-e^{-i\gamma_3}}{2}) \right] \\\nonumber
    {} & + s_{\alpha}s_{\phi}\left[ (\frac{1}{\sqrt{2}})^{N-1}\frac{-i}{\sqrt{2}}(\frac{e^{i(\gamma_4-\gamma_1)}-e^{-i(\gamma_4-\gamma_1)}}{2}+\frac{e^{i(\gamma_5-\gamma_1)}-e^{-i(\gamma_5-\gamma_1)}}{2})\right.\\\nonumber
    {} &\hspace{5cm} +\left.(\frac{i}{\sqrt{2}})^{N-1}(\frac{-e^{i\gamma_5}+e^{-i\gamma_4}}{2\sqrt{2}})+(\frac{-i}{\sqrt{2}})^{N-1}(\frac{e^{i\gamma_4}-e^{-i\gamma_5}}{2\sqrt{2}})\right] \\\nonumber
    {} &  + s_{\theta}s_{\phi}\left[ (\frac{1+i}{2})^N(\frac{-e^{i(\gamma_4-\gamma_3)}-e^{i(\gamma_5-\gamma_2)}+e^{-i(\gamma_4-\gamma_2)}+e^{-i(\gamma_5-\gamma_3)}}{2})\right.\\\nonumber
    {} &\hspace{5cm}+\left.\left.(\frac{1-i}{2})^N(\frac{e^{i(\gamma_4-\gamma_2)}+e^{i(\gamma_5-\gamma_3)}-e^{-i(\gamma_4-\gamma_3)}-e^{-i(\gamma_5-\gamma_2)}}{2}) \right]\right\},
\end{align}

and with $r_{x0}=\bra{\Phi_o}\bm{n}_x\cdot\bm{\sigma}\ket{\Phi_o}$ given as
\begin{align}
    r_{\alpha 0} &=\bra{\Phi_o}\bm{n}_\alpha\cdot\bm{\sigma}\ket{\Phi_o} \\\nonumber &=\frac{1}{M^2}\left\{s_{\alpha}s_{\theta} \left[ (\frac{1}{\sqrt{2}})^N(\frac{e^{i\gamma_2}+e^{-i\gamma_2}}{2}-\frac{e^{i(\gamma_2-\gamma_1)}+e^{-i(\gamma_2-\gamma_1)}}{2}-\frac{e^{i(\gamma_3-\gamma_1)}+e^{-i(\gamma_3-\gamma_1)}}{2})\right.\right.\\\nonumber
    {}&\hspace{5cm}+\left.(\frac{-1}{\sqrt{2}})^N(\frac{e^{i\gamma_3}+e^{-i\gamma_3}}{2}) \right] \\\nonumber
    {} & + s_{\alpha}s_{\phi}\left[ -(\frac{1}{\sqrt{2}})^N(\frac{e^{i(\gamma_4-\gamma_1)}+e^{-i(\gamma_4-\gamma_1)}}{2}+\frac{e^{i(\gamma_5-\gamma_1)}+e^{-i(\gamma_5-\gamma_1)}}{2})\right.\\\nonumber
    {} &\hspace{5cm} +\left.(\frac{i}{\sqrt{2}})^N(\frac{e^{i\gamma_5}+e^{-i\gamma_4}}{2})+(\frac{-i}{\sqrt{2}})^N(\frac{e^{i\gamma_4}+e^{-i\gamma_5}}{2})\right] \\\nonumber
    {} &  + s_{\theta}s_{\phi}\left[ -(\frac{1+i}{2})^{N-2}\frac{e^{i(\gamma_4-\gamma_3)}+e^{-i(\gamma_4-\gamma_2)}+e^{i(\gamma_5-\gamma_2)}+e^{-i(\gamma_5-\gamma_3)}}{4}\right.\\\nonumber
    {} &\hspace{5cm}\left.\left.-(\frac{1-i}{2})^{N-2}\frac{e^{i(\gamma_4-\gamma_2)}+e^{-i(\gamma_4-\gamma_3)}+e^{i(\gamma_5-\gamma_3)}+e^{-i(\gamma_5-\gamma_2)}}{4} \right]\right\},
\end{align}

\begin{align}
    r_{\theta 0} &=\bra{\Phi_o}\bm{n}_\theta\cdot\bm{\sigma}\ket{\Phi_o} \\\nonumber &=\frac{1}{M^2}\left\{s_{\alpha}s_{\theta} \left[ (\frac{1}{\sqrt{2}})^N(\frac{e^{i\gamma_2}+e^{-i\gamma_2}}{2}+\frac{e^{i(\gamma_2-\gamma_1)}+e^{-i(\gamma_2-\gamma_1)}}{2}-\frac{e^{i(\gamma_3-\gamma_1)}+e^{-i(\gamma_3-\gamma_1)}}{2})\right.\right.\\\nonumber
    {}&\hspace{5cm}-\left.(\frac{-1}{\sqrt{2}})^N(\frac{e^{i\gamma_3}+e^{-i\gamma_3}}{2}) \right] \\\nonumber
    {} & + s_{\alpha}s_{\phi}\left[ (\frac{1}{\sqrt{2}})^{N-1}\frac{i}{\sqrt{2}}(\frac{-e^{i(\gamma_4-\gamma_1)}+e^{-i(\gamma_4-\gamma_1)}+e^{i(\gamma_5-\gamma_1)}-e^{-i(\gamma_5-\gamma_1)}}{2})\right.\\\nonumber
    {} &\hspace{5cm} +\left.(\frac{i}{\sqrt{2}})^{N-1}(\frac{e^{i\gamma_5}+e^{-i\gamma_4}}{2\sqrt{2}})+(\frac{-i}{\sqrt{2}})^{N-1}(\frac{e^{i\gamma_4}+e^{-i\gamma_5}}{2\sqrt{2}})\right] \\\nonumber
    {} &  + s_{\theta}s_{\phi}\left[ (\frac{1+i}{2})^N\frac{-e^{i(\gamma_4-\gamma_3)}+e^{i(\gamma_5-\gamma_2)}+e^{-i(\gamma_4-\gamma_2)}-e^{-i(\gamma_5-\gamma_3)}}{2}\right.\\\nonumber
    {} &\hspace{5cm}+\left.\left.(\frac{1-i}{2})^N\frac{e^{i(\gamma_4-\gamma_2)}-e^{i(\gamma_5-\gamma_3)}-e^{-i(\gamma_4-\gamma_3)}+e^{-i(\gamma_5-\gamma_2)}}{2} \right]\right\},
\end{align}

\begin{align}
    r_{\phi 0} &=\bra{\Phi_o}\bm{n}_\phi\cdot\bm{\sigma}\ket{\Phi_o} \\\nonumber &=\frac{1}{M^2}\left\{s_{\alpha}s_{\theta} \left[ (\frac{1}{\sqrt{2}})^{N-1}\frac{i}{\sqrt{2}}(\frac{-e^{i\gamma_2}+e^{-i\gamma_2}}{2}+\frac{e^{i(\gamma_2-\gamma_1)}-e^{-i(\gamma_2-\gamma_1)}}{2}+\frac{-e^{i(\gamma_3-\gamma_1)}+e^{-i(\gamma_3-\gamma_1)}}{2})\right.\right.\\\nonumber
    {}&\hspace{5cm}+\left.(\frac{-1}{\sqrt{2}})^{N-1}\frac{i}{\sqrt{2}}(\frac{-e^{i\gamma_3}+e^{-i\gamma_3}}{2}) \right] \\\nonumber
    {} & + s_{\alpha}s_{\phi}\left[ (\frac{1}{\sqrt{2}})^N(\frac{e^{i(\gamma_4-\gamma_1)}+e^{-i(\gamma_4-\gamma_1)}}{2}-\frac{e^{i(\gamma_5-\gamma_1)}+e^{-i(\gamma_5-\gamma_1)}}{2})\right.\\\nonumber
    {} &\hspace{5cm} +\left.(\frac{i}{\sqrt{2}})^N(\frac{-e^{i\gamma_5}+e^{-i\gamma_4}}{2})+(\frac{-i}{\sqrt{2}})^N(\frac{e^{i\gamma_4}-e^{-i\gamma_5}}{2})\right] \\\nonumber
    {} &  + s_{\theta}s_{\phi}\left[ (\frac{1+i}{2})^N\frac{e^{i(\gamma_4-\gamma_3)}-e^{i(\gamma_5-\gamma_2)}+e^{-i(\gamma_4-\gamma_2)}-e^{-i(\gamma_5-\gamma_3)}}{2}\right.\\\nonumber
    {} &\hspace{5cm}+\left.\left.(\frac{1-i}{2})^N\frac{e^{i(\gamma_4-\gamma_2)}-e^{i(\gamma_5-\gamma_3)}+e^{-i(\gamma_4-\gamma_3)}-e^{-i(\gamma_5-\gamma_2)}}{2} \right]\right\}.
\end{align}

Then the quantum Fisher information matrix can be easily computed using the reduced two-spin state and the reduced single spin state, given by $J=4NJ_1+4N(N-1)J_2-4N^2J_3$, where
\begin{eqnarray}
J_1&=\left(
\begin{array}{ccc}
1 & 0 & 0\\
0 & \sin^2\alpha & 0\\
0 & 0 & \sin^2\alpha\sin^2\theta\\
\end{array}
\right),\\
J_2&=\left(
\begin{array}{ccc}
r_{\alpha\alpha} &  r_{\alpha\theta}\sin\alpha & r_{\alpha\phi}\sin\alpha\sin\theta \\
 r_{\alpha\theta}\sin\alpha &  r_{\theta\theta}\sin^2\alpha &  r_{\theta\phi}\sin^2\alpha\sin\theta\\
 r_{\alpha\phi}\sin\alpha\sin\theta &  r_{\theta\phi}\sin^2\alpha\sin\theta &  r_{\phi\phi}\sin^2\alpha\sin^2\theta\\
\end{array}
\right),\\
J_3&=\left(
\begin{array}{ccc}
r_{\alpha0}^2 &  r_{\alpha 0}r_{\theta 0}\sin\alpha &  r_{\alpha 0}r_{\phi 0}\sin\alpha\sin\theta\\
 r_{\alpha 0}r_{\theta 0}\sin\alpha &  r_{\theta 0}^2\sin^2\alpha &  r_{\theta 0}r_{\phi 0}\sin^2\alpha\sin\theta\\
 r_{\alpha 0}r_{\phi 0}\sin\alpha\sin\theta &  r_{\theta 0}r_{\phi 0}\sin^2\alpha\sin\theta &  r_{\phi 0}^2\sin^2\alpha\sin^2\theta\\
\end{array}
\right),\\
\end{eqnarray}

When $N=4n$, let $\gamma_i=0$ with $i=1,2,\cdots, 5$, we have $\rho^{[2]}=\frac{1}{4}[I+\sum_{x\in\{\alpha,\theta,\phi\}} r_{xx}\sigma_x\otimes\sigma_x]$ with
\begin{eqnarray}
\aligned
r_{\alpha\alpha}&=\frac{1}{M^2}[s^2_{\alpha}+(\frac{1}{\sqrt{2}})^{N-4}(s_{\alpha}s_{\theta}+s_{\alpha}s_{\phi}-(-1)^{N/4}s_{\theta}s_{\phi})],\\
r_{\theta\theta}&=\frac{1}{M^2}[s^2_{\theta}+(\frac{1}{\sqrt{2}})^{N-4}(s_{\alpha}s_{\theta}-s_{\alpha}s_{\phi}+(-1)^{N/4}s_{\theta}s_{\phi})],\\
r_{\phi\phi}&=\frac{1}{M^2}[s^2_{\phi}+(\frac{1}{\sqrt{2}})^{N-4}(-s_{\alpha}s_{\theta}+s_{\alpha}s_{\phi}+(-1)^{N/4}s_{\theta}s_{\phi})],\\
\endaligned
\end{eqnarray}
here $M^2=s^2_{\alpha}+s^2_{\theta}+s^2_{\phi}+(\frac{1}{\sqrt{2}})^{N-4}(s_{\alpha}s_{\theta}+s_{\alpha}s_{\phi}+(-1)^{N/4}s_{\theta}s_{\phi})]$. When $N\rightarrow \infty$, we can achieve the ultimate lower bound by taking $s_{\alpha}=\sqrt{\tilde{r}_{\alpha\alpha}}$, $s_{\theta}=\sqrt{\tilde{r}_{\theta\theta}}$ and $s_{\phi}=\sqrt{\tilde{r}_{\phi\phi}}$. For finite N, we numerically optimize the coefficients to get the best precision.

\section{Measurement saturating the QCRB \label{optmeasurement}}

For the probe state without the ancillary system, $\left|\Phi_{o}\right\rangle=\sqrt{\tilde{r}_{\alpha \alpha}}\left|\Phi_{\alpha}\right\rangle+\sqrt{\tilde{r}_{\theta \theta}}\left|\Phi_{\theta}\right\rangle+\sqrt{\tilde{r}_{\phi \phi}}\left|\Phi_{\phi}\right\rangle$, we first check the weak commutativity condition, which is $\mathrm{Im}[ \left<\partial_x\Phi(\alpha,\theta,\phi)|\partial_y\Phi(\alpha,\theta,\phi)\right>]=0$ for all $x,y\in\{\alpha,\theta,\phi\}$, here $\ket{\Phi(\alpha,\theta,\phi)}=U_s\ket{\Phi_o}$ is the output state with $U_s=e^{-\rmi \alpha\bm{n}\cdot\bm{\sigma}}\ket{\Phi_o}$. $\ket{\partial_x\Phi(\alpha,\theta,\phi)}=-\rmi U_sH_x^{(N)}\ket{\Phi_o}$ $\forall x\in\{\alpha,\theta,\phi\}$.
It is easy to compute
\begin{equation}
  \begin{aligned}
    \mathrm{Im} \left[\left<\partial_x\Phi(\alpha,\theta,\phi)|\partial_y\Phi(\alpha,\theta,\phi)\right>\right]&=\mathrm{Im}\left[ \bra{\Phi_{o}}H_x^{(N)}H_y^{(N)}\ket{\Phi_{o}}\right]\\
    &= \mathrm{Im} \left[ \sum_{j=0}^{N-1}\bra{\Phi_{o}}H_x^{[j]}H_y^{[j]}\ket{\Phi_{o}}+\sum_{j\neq k}c_x c_y \bra{\Phi_{o}}H_x^{[j]}H_y^{[k]}\ket{\Phi_{o}} \right]\\
    &=\sum_{k=0}^{N-1}c_x c_y \epsilon_{xyz}r_{z0}^{(k)},
  \end{aligned}
\end{equation}
here $x,y,z$ represent three different parameters in $\{\theta,\phi,\alpha\}$ and $\epsilon_{xyz}$ is the Levi-Civita symbol under the label $(\theta,\phi,\alpha)=(1,2,3)$, and for the last equality we have used the fact that  $H_x^{[j]}H_y^{[k]}=i\epsilon_{xyz}H_z^{[k]}$, $r_{z0}^{(k)}=\bra{\Phi_{o}}H_z^{(k)}\ket{\Phi_{o}}$ and $r_{xy}^{(j,k)}=\bra{\Phi_{o}}H_x^{[j]}H_y^{[k]}\ket{\Phi_{o}}$ are real numbers.
The weak commutativity condition is thus equivalent to $r_{z0}^{(k)}=\tr[\rho^{(k)}H_z^{(k)}]=0$ for all $z\in\{\alpha,\theta,\phi\}$(note that the state is permutation invariant so $r_{z0}^{(k)}$ is the same for all $k$).
For $N=4n$ with $n\in\mathbb{N}$ or for $N\rightarrow\infty$, the condition holds as the reduced single spin state is given by $\rho^{(k)}=\frac{I^{(k)}}{2}$. Thus there exists a set of POVM which saturates the QCRB.

One can construct such a set of POVM, denoted as $\{\Pi_k\}$, to saturate the QCRB. Following the studies in \cite{PhysRevLett.111.070403,PhysRevLett.119.130504}, a set of POVM can be consisted with the projective measurements onto the space spanned by the state and the partial derivatives, since the QFIM only depends on the state and its partial derivatives, i.e., the measurement can be chosen as  $\{\Pi_k=\ket{\xi_k}\bra{\xi_k}\}_{k=0}^3$ with
\begin{equation}\label{eq.optm}
  \begin{aligned}
    \ket{\xi_0}&=\ket{\Phi(\alpha,\theta,\phi)}=U_s\ket{\Phi_o},\\ \ket{\xi_1}&=\frac{\ket{\partial_\alpha\Phi(\alpha,\theta,\phi)}}{\sqrt{\left<\partial_\alpha\Phi(\alpha,\theta,\phi)|\partial_\alpha\Phi(\alpha,\theta,\phi)\right>}}=-\rmi U_s\bm{\sigma}_\alpha^{(N)}\ket{\Phi_o},\\
    \ket{\xi_2}&=\frac{\ket{\partial_\theta\Phi(\alpha,\theta,\phi)}}{\sqrt{\left<\partial_\theta\Phi(\alpha,\theta,\phi)|\partial_\theta\Phi(\alpha,\theta,\phi)\right>}}=-\rmi U_s\bm{\sigma}_\theta^{(N)}\ket{\Phi_o},\\
    \ket{\xi_3}&=\frac{\ket{\partial_\phi\Phi(\alpha,\theta,\phi)}}{\sqrt{\left<\partial_\phi\Phi(\alpha,\theta,\phi)|\partial_\phi\Phi(\alpha,\theta,\phi)\right>}}=-\rmi U_s\bm{\sigma}_\phi^{(N)}\ket{\Phi_o},\\
  \end{aligned}
\end{equation}
with an additional element $\Pi_4=I-\sum_{k=0}^3\Pi_k$ that accounts for the normalization,
here $\bm{\sigma}_x^{(N)}=\sum_{j=0}^{N-1}\bm{n}_x\cdot\bm{\sigma}^{[j]}$ for $x=\alpha,\theta,\phi$ and $\bm{n}_x\cdot\bm{\sigma}$ is defined in the main text with an additional superscript $[j]$ indicating that the operator only acts on the $j$-th spin.
It can be directly checked that vectors in Eq.(\ref{eq.optm}) are orthogonal to each other when $N=4n$ or $N\rightarrow \infty$.
For example,
\begin{equation}
  \begin{aligned}
    \braket{\xi_1}{\xi_2}=&\bra{\Phi_o}\bm{\sigma}_\alpha^{(N)}\bm{\sigma}_\theta^{(N)}\ket{\Phi_o}=\bra{\Phi_o}\left(\sum_{j=0}^{N-1}\bm{n}_\alpha\cdot\bm{\sigma}^{[j]}\right)\left(\sum_{k=0}^{N-1}\bm{n}_\theta\cdot\bm{\sigma}^{[k]}\right)\ket{\Phi_o}\\
    =& N\bra{\Phi_o}(\bm{n}_\alpha\cdot\bm{\sigma}^{[1]})(\bm{n}_\theta\cdot\bm{\sigma}^{[1]})\ket{\Phi_o}+N(N-1)\bra{\Phi_o}(\bm{n}_\alpha\cdot\bm{\sigma}^{[1]})\otimes(\bm{n}_\theta\cdot\bm{\sigma}^{[1]})\ket{\Phi_o}\\
    =& iNr_{\phi 0}+N(N-1)r_{\alpha\theta}=0
  \end{aligned}
\end{equation}
since $r_{\phi 0}=r_{\alpha\theta}=0$ when $N=4n$ or $N\rightarrow \infty$.
The probability distribution of the measurement outcome under $\{\Pi_k\}_{k=0}^4$ is simply $p_0=1$ and $p_k=0$ for $k=1,2,3,4$.
But the classical Fisher information matrix (CFIM) also depends on the derivative of $p_k$ with respect to $\alpha$, $\theta$ and $\phi$, i.e.,
\begin{equation}
  \begin{aligned}
    I_{x,y\in\{\alpha,\theta,\phi\}}&=\sum_{k=0}^4 \frac{(\partial_x p_k)(\partial_y p_k)}{p_k}=\sum_{k=0}^4 \frac{(\partial_x \bra{\Phi(\alpha,\theta,\phi)}\Pi_k \ket{\Phi(\alpha,\theta,\phi)})(\partial_y \bra{\Phi(\alpha,\theta,\phi)}\Pi_k \ket{\Phi(\alpha,\theta,\phi)})}{\bra{\Phi(\alpha,\theta,\phi)} \Pi_k \ket{\Phi(\alpha,\theta,\phi)}}\\
    &= \sum_{k=0}^4 \frac{4\mathrm{Re}(\bra{\partial_x\Phi(\alpha,\theta,\phi)}\Pi_k \ket{\Phi(\alpha,\theta,\phi)})\mathrm{Re}( \bra{\Phi(\alpha,\theta,\phi)}\Pi_k \ket{\partial_y\Phi(\alpha,\theta,\phi)})}{\bra{\Phi(\alpha,\theta,\phi)} \Pi_k \ket{\Phi(\alpha,\theta,\phi)}}.
  \end{aligned}
\end{equation}
For $k=0$, i.e. $\Pi_0=\ket{\Phi(\alpha,\theta,\phi)}\bra{\Phi(\alpha,\theta,\phi)}$, the term is of the form $\frac{0}{1}$ since $\mathrm{Re}[\left<\partial_x\Phi(\alpha,\theta,\phi)| \Phi(\alpha,\theta,\phi)\right>]=\partial_x \left<\Phi(\alpha,\theta,\phi)| \Phi(\alpha,\theta,\phi)\right>=0$.
The other terms have the form $\frac{0}{0}$ for $k=1,2,3,4$, which need to be calculated via the limit. We can evaluate $I_{x,y}$ when the paremeters $\alpha$, $\theta$ or $\phi$ are displaced by a small disturbance $\delta\alpha$, $\delta\theta$ or $\delta\phi$, respectively. The disturbance can be taken as arbitrary small values since the elements of CFIM do not dependent on the actual value of the disturbance.
By replacing $\ket{\Phi(\alpha,\theta,\phi)}$ with $\ket{\Phi(\alpha,\theta,\phi)}+\sum_{l=1}^3\delta x_l  \ket{\partial_{x_l} \Phi(\alpha,\theta,\phi)}$, where $x_1=\alpha$, $x_2=\theta$, $x_3=\phi$ and use the fact that the vectors in Eq.(\ref{eq.optm}) are orthogonal to each other, one can directly find that
\begin{equation}
  \begin{aligned}
    I_{x,y\in\{\alpha,\theta,\phi\}}&= \sum_{k=1}^4 \frac{4\delta x_k^2\mathrm{Re}(\bra{\partial_x\Phi(\alpha,\theta,\phi)}\Pi_k \ket{\partial_{x_k}\Phi(\alpha,\theta,\phi)})\mathrm{Re}( \bra{\partial_{x_k}\Phi(\alpha,\theta,\phi)}\Pi_k \ket{\partial_y\Phi(\alpha,\theta,\phi)})}{\delta x_k^2\bra{\partial_{x_k}\Phi(\alpha,\theta,\phi)} \Pi_k \ket{\partial_{x_k}\Phi(\alpha,\theta,\phi)}}\\
    &= \sum_{k=1}^4 \frac{4\mathrm{Re}(\bra{\partial_x\Phi(\alpha,\theta,\phi)}\Pi_k \ket{\partial_{x_k}\Phi(\alpha,\theta,\phi)})\delta_{y,x_k}\mathrm{Re}( \bra{\partial_{x_k}\Phi(\alpha,\theta,\phi)}\Pi_k \ket{\partial_y\Phi(\alpha,\theta,\phi)})}{\bra{\partial_{x_k}\Phi(\alpha,\theta,\phi)} \Pi_k \ket{\partial_{x_k}\Phi(\alpha,\theta,\phi)}}\\
    &=\sum_{k=1}^4 4\mathrm{Re}(\bra{\partial_x \Phi(\alpha,\theta,\phi)}\Pi_k \ket{\partial_y \Phi(\alpha,\theta,\phi)})\\
    &= 4\mathrm{Re}(\bra{\partial_x \Phi(\alpha,\theta,\phi)}(I-\Pi_0) \ket{\partial_y \Phi(\alpha,\theta,\phi)})\\
    &= 4\mathrm{Re}(\braket{\partial_x \Phi(\alpha,\theta,\phi)}{\partial_y \Phi(\alpha,\theta,\phi)}- \braket{\partial_x \Phi(\alpha,\theta,\phi)}{\Phi(\alpha,\theta,\phi)} \braket{\Phi(\alpha,\theta,\phi)}{\partial_y \Phi(\alpha,\theta,\phi)})\\
    &=J_{x,y\in\{\alpha,\theta,\phi\}},
  \end{aligned}
\end{equation}
i.e. the QCRB is saturated with the given measurement basis.

For the case with the ancillary qutrit, the probe state is $ | \Psi_{SA} \rangle=s_\alpha | \Phi_{\alpha} \rangle\otimes\ket{0}+s_\theta | \Phi_{\theta} \rangle\otimes\ket{1}+s_\phi| \Phi_{\phi} \rangle\otimes\ket{2}$.
Denote $|\Psi_{SA}(\alpha,\theta,\phi)\rangle$ as the output state, it can be verified that the weak commutativity condition, $\mathrm{Im}[\left<\partial_x\Psi_{SA}(\alpha,\theta,\phi)|\partial_y\Psi_{SA}(\alpha,\theta,\phi)\right>]=0$ for all $x,y\in\{\alpha,\theta,\phi\}$, is satisfied for any $N$ as with the ancillary qutrit the reduced single spin state is always $\frac{I}{2}$.
Following the identical procedure, one can obtain the optimal POVM saturating the QCRB as $\{\prod_k=\ket{\xi_k}\bra{\xi_k}\}_{k=0}^3$, where
\begin{equation}\label{eq:appoptm}
  \begin{aligned}
    \ket{\xi_0}&=U_s\otimes I_A\ket{\Psi_{SA}},\\ \ket{\xi_1}&=-\rmi U_s\bm{\sigma}_\alpha^{(N)}\otimes I_A\ket{\Psi_{SA}},\\ \ket{\xi_2}&=-\rmi U_s\bm{\sigma}_\theta^{(N)}\otimes I_A\ket{\Psi_{SA}},\\
    \ket{\xi_3}&=-\rmi U_s\bm{\sigma}_\phi^{(N)}\otimes I_A\ket{\Psi_{SA}},\\
  \end{aligned}
\end{equation}
together with an additional element $\Pi_4=I-\sum_{k=0}^3\Pi_k$ for the normalization.

\section{Ultimate precision with general spin-S}

In this section we derive the ultimate precision for the estimation of the magnetic field using general spin-S, where the Hamiltonian can be written as $H=B\vec{n}\cdot \vec{S}=B_1S_1+B_2S_2+B_3S_3$, where $B$ is the magnitude of the magnetic field and $\bm{n}=(\sin\theta\cos\phi , \sin\theta\sin\phi, \cos\theta)$ is the direction of the magnetic field, $\vec{S}=(S_1,S_2,S_3)$ is the spin vectors for spin-S which satisfies the commutation relations as
\begin{equation}\label{eq:appcommute}
  [S_1,S_2]=\rmi S_3,\quad [S_2,S_3]=\rmi S_1,\quad [S_3,S_1]=\rmi S_2.
\end{equation}
For spin-1/2 particles, these spin vectors can be written in terms of Pauli matrices as $\bm{S}=\bm{\sigma}/2$.
For general spin-$S$ these operators are Hermitian matrices of dimension $2S+1$. In a basis called the Zeeman basis, denoted as $\ket{S,m}$ or in short $\ket{m}$, with $m=-S,...,S$, the entries of the spin operators are given by
\begin{equation}\label{eq:appspinmatrix}
  \begin{aligned}
    \bra{m'}S_1\ket{m}&=(\delta_{m',m+1}+\delta_{m'+1,m})\dfrac{1}{2}\sqrt{S(S+1)-m'm},\\
    \bra{m'}S_2\ket{m}&=(\delta_{m',m+1}-\delta_{m'+1,m})\dfrac{1}{2\rmi}\sqrt{S(S+1)-m'm},\\
    \bra{m'}S_3\ket{m}&=\delta_{m',m}m.
  \end{aligned}
\end{equation}
We note that for $S>\frac{1}{2}$, $S_j^2\neq I$, $\forall j\in \{1,2,3\}$. This is different from the Pauli matrices. The analysis for spin-1/2 can not be directly used for spin-S.

After an evolution time $t$, the dynamics generates a unitary operator as $U_s=e^{-\rmi \alpha\bm{n}\cdot\bm{S}}$ with $\alpha=Bt$.
We now derive the generators for the three parameters in this case. During the derivation we will make use of the following formulas: for two general three-dimensional vectors $\bm{a}$ and $\bm{b}$,
\begin{equation}\label{eq:approtation}
  e^{-\rmi\theta\bm{a}\cdot\bm{S}}\left(\bm{b}\cdot\bm{S}\right)e^{\rmi\theta\bm{a}\cdot\bm{S}}=\left[\sin\theta\left(\bm{a}\times\bm{b}\right)+\left(1-\cos\theta\right)\left(\bm{a}\cdot\bm{b}\right)\bm{a}+\cos\theta\bm{b}\right]\cdot\bm{S}.
\end{equation}
and\cite{Wilcox1967}
\begin{equation}\label{eq:dU}
  \frac{\partial_g e^{-iH(g)}}{\partial g} =-i\int_0^1e^{-isH(g)}\frac{\partial_gH(g)}{\partial g} e^{i(s-1)H(g)}ds.
\end{equation}
Now for $x\in\{\alpha,\theta,\phi\}$, the corresponding generator can be obtained as
\begin{equation}
  \begin{aligned}\label{eq:appgenerator}
    H_x&=\rmi U_s^\dagger(\partial_x U_s)=e^{\rmi \alpha\bm{n}\cdot\bm{S}}\int_0^1 e^{-\rmi s\alpha\bm{n}\cdot\bm{S}}\partial_x (\alpha\bm{n}\cdot\bm{S}) e^{\rmi (s-1)\alpha\bm{n}\cdot\bm{S}}ds\\
    &= \int_{-1}^{0} e^{-\rmi s\alpha\bm{n}\cdot\bm{S}}\left[\partial_x (\alpha\bm{n})\cdot\bm{S}\right] e^{\rmi s\alpha\bm{n}\cdot\bm{S}}ds.
  \end{aligned}
\end{equation}
Using Eq.(\ref{eq:approtation}), one can then easily obtain the three generators as
\begin{equation}
  \begin{aligned}
    H_\alpha=&\bm{n}\cdot\bm{S}= c_\alpha(\bm{n_\alpha}\cdot\bm{S})= c_\alpha{S_\alpha},\\
    H_\theta=& \left[\sin\alpha\bm{n}_1-(1-\cos\alpha)(\bm{n}\times\bm{n}_1)\right]\cdot\bm{S}\\
    =& 2\sin\frac{\alpha}{2}\left(\cos\frac{\alpha}{2}\bm{n}_1-\sin\frac{\alpha}{2}\bm{n}_2\right)\cdot\bm{S}= c_\theta(\bm{n_\theta}\cdot\bm{S})= c_\theta S_\theta,\\
    H_\phi=& \sin\theta\left[\sin\alpha(\bm{n}\times\bm{n}_1)+(1-\cos\alpha)\bm{n}_1\right]\cdot\bm{S}\\
    =& 2\sin\frac{\alpha}{2}\sin\theta\left(\sin\frac{\alpha}{2}\bm{n}_1+\cos\frac{\alpha}{2}\bm{n}_2\right)\cdot\bm{S}= c_\phi(\bm{n_\phi}\cdot\bm{S})= c_\phi S_\phi,
  \end{aligned}
\end{equation}
with $c_\alpha=1$, $S_\alpha=\bm{n}_\alpha\cdot\bm{S}$, $\bm{n}_\alpha=\bm{n}=(\sin\theta\cos\phi,\sin\theta\sin\phi,\cos\theta)$, $c_\theta=2\sin\frac{\alpha}{2}$, $S_\theta=\bm{n}_\theta\cdot\bm{S}$, $\bm{n}_\theta=\cos\frac{\alpha}{2}\bm{n}_1-\sin\frac{\alpha}{2}\bm{n}_2$, $c_\phi=2\sin\frac{\alpha}{2}\sin\theta$, $S_\phi=\bm{n}_\phi\cdot\bm{S}$, $\bm{n}_\phi=\sin\frac{\alpha}{2}\bm{n}_1+\cos\frac{\alpha}{2}\bm{n}_2$, here
$\bm{n_1}=\partial_\theta\bm{n}=(\cos\theta\cos\phi,\cos\theta\sin\phi,-\sin\theta)$, $\bm{n_2}=\bm{n}\times\bm{n_1}=(-\sin\phi,\cos\phi,0)$.
It is also worthy noting that there exists a unitary rotation $U_r=\exp(\rmi\frac{\alpha}{2}\bm{n}\cdot\bm{S})\exp(-\rmi\phi S_3)\exp(-\rmi\theta S_2)$ such that
\begin{equation}
  S_\alpha=U_r S_3 U_r^\dagger, S_\theta=U_r S_1 U_r^\dagger, S_\phi=U_r S_2 U_r^\dagger.
\end{equation}


With $N$ spins interacting with the magnetic field and an ancillary system, the generator for each $x\in\{\alpha,\theta,\phi\}$ is
\begin{equation}
  H_x^{(N)}=\sum_{k=0}^{N-1}H_x^{[k]},
\end{equation}
where $H_x^{[k]}=I\otimes\cdots\otimes I\otimes H_x\otimes I\cdots\otimes I\otimes I_A$ denotes the generator on the $k$th spin.
The variance of $H_x^{(N)}$ is given by
\begin{equation}
  \left< \Delta\left[H_x^{(N)}\right]^2 \right>=\left< \left[H_x^{(N)}\right]^2 \right> - \left< H_x^{(N)}\right>^2,
\end{equation}
where the first term can be expanded as
\begin{equation}
  \begin{aligned}
    \left< \left[H_x^{(N)}\right]^2 \right>=& \sum_{k=0}^{N-1} \left< \left(H_x^{[k]}\right)^2 \right>+\sum_{j\neq k} \left< H_x^{[j]}H_x^{[k]} \right>\\
    =& c_x^2 \left[ \sum_{k=0}^{N-1}\tr\left( \rho^{(k)}S_x^2 \right)+\sum_{j\neq k}\tr\left( \rho^{(j,k)}S_x\otimes S_x \right) \right]\\
    =& c_x^2 \left[ \sum_{k=0}^{N-1}r^{(k)}_{xx}+\sum_{j\neq k}r^{(j,k)}_{xx} \right],
  \end{aligned}
\end{equation}
and the second term as
\begin{equation}
  \left< H_x^{(N)}\right>^2=c_x^2\left[ \sum_{k=0}^{N-1}\tr\left( \rho^{(k)}S_x \right) \right]^2= c_x^2\left[ \sum_{k=0}^{N-1}r^{(k)}_x \right]^2,
\end{equation}
here $r^{(k)}_{xx}=\tr\left( \rho^{(k)}S_x^2 \right)\le S^2$, $r^{(j,k)}_{xx}=\tr\left( \rho^{(j,k)}S_x\otimes S_x \right)\le S^2$ and $r^{(k)}_x=\tr\left( \rho^{(k)}S_x \right)$ since the largest and smallest eigenvalues of $S_x$ are $\pm S$.

If there is only one parameter in $\{\alpha,\theta,\phi\}$ to be estimated, one can choose the probe state as $\ket{\Phi_x}=\frac{1}{\sqrt{2}}\left(\ket{S^+_x}^{\otimes N}+\ket{S^-_x}^{\otimes N}\right)$, where $S_x\ket{S^{\pm}_x}=\pm S\ket{S^{\pm}_x}$.
The reduced single-spin states and two-spin states are $\rho^{(k)}=\frac{1}{2}(\ket{S^+_x}\bra{S^+_x}^{(k)}+\ket{S^-_x}\bra{S^-_x}^{(k)})$ and $\rho^{(j,k)}=\frac{1}{2}\left[\ket{S^+_x}\bra{S^+_x}^{(j)}\otimes \ket{S^+_x}\bra{S^+_x}^{(k)}+\ket{S^-_x}\bra{S^-_x}^{(j)}\otimes\ket{S^-_x}\bra{S^-_x}^{(k)}\right]$,
then $r^{(k)}_{xx}=r^{(j,k)}_{xx}=S^2$, $r^{(k)}_x=0$, $\forall j,k$, and $\left< \Delta\left[H_x^{(N)}\right]^2 \right>=N^2S^2c_x^2$.
This achieves the best precision for a single parameter as
\begin{equation}
  \delta\hat{x}^2\ge\frac{1}{4N^2S^2c_x^2},\quad x\in\{\alpha,\theta,\phi\}.
\end{equation}
For the estimation of all three parameters, however, the tradeoff is also unavoidable. We now characterize the minimal tradeoff among the precision of the three parameters for general spin-S.

For each parameter $x\in\{\alpha,\theta,\phi\}$, we have $\delta \hat{x}^{2} \geq \frac{1}{4\langle\Delta [H_{x}^{(N)}]^{2}\rangle}$. With a similar procedure we can get
\begin{equation}\label{eq:appbound1}
  \begin{aligned}
    w_\alpha\delta \hat{\alpha}^{2}+w_\theta\delta \hat{\theta}^{2}+w_\phi\delta \hat{\phi}^{2} &\geq \frac{1}{4}\sum_{x\in\{\alpha,\theta,\phi\}} \frac{w_x}{\langle\Delta [H_{x}^{(N)}]^{2}\rangle}\\
    &=\frac{1}{4}\sum_{x\in\{\alpha,\theta,\phi\}} \frac{w_x/c_x^2}{\left[ \sum_{k=0}^{N-1}r^{(k)}_{xx}+\sum_{j\neq k}r^{(j,k)}_{xx} \right]-\left[ \sum_{k=0}^{N-1}r^{(k)}_x \right]^2}\\
    &\geq \frac{1}{4}\sum_{x\in\{\alpha,\theta,\phi\}} \frac{w_x/c_x^2}{\left[ \sum_{k=0}^{N-1}r^{(k)}_{xx}+\sum_{j\neq k}r^{(j,k)}_{xx} \right]}\\
    &\geq \frac{1}{4}\frac{\left(\sum_x \sqrt{w_x}/|c_x|\right)^2} {\left[ \sum_{k=0}^{N-1}\sum_x r^{(k)}_{xx}+\sum_{j\neq k}\sum_x r^{(j,k)}_{xx} \right]},
  \end{aligned}
\end{equation}
where the last inequality is obtained from the Cauchy-Schwarz inequality.
To get an explicit lower bound, we need to characterize the constraints on various terms in the denominator.
First we have
\begin{equation}\label{eq:apprxxk}
  \sum_{x\in\{\alpha,\theta,\phi\}} r^{(k)}_{xx}=\sum_{x\in\{\alpha,\theta,\phi\}} \tr\left( \rho^{(k)}S_x^2 \right)=S(S+1) \tr\left( \rho^{(k)}\mathbb{I}_{2S+1} \right)=S(S+1),
\end{equation}
where we used the fact that $\sum_{x\in\{\alpha,\theta,\phi\}}S_x^2=U_r(S_1^2+S_2^2+S_3^2)U_r^\dagger=S(S+1)\mathbb{I}_{2S+1}$ with $\mathbb{I}_{2S+1}$ as the identity operator of dimension $2S+1$.
Second
\begin{equation}
  \sum_{x\in\{\alpha,\theta,\phi\}} r^{(j,k)}_{xx}=\sum_{x\in\{\alpha,\theta,\phi\}} \tr\left( \rho^{(j,k)}S_x\otimes S_x \right)=\tr\left( \tilde{\rho}^{(j,k)}\sum_{i\in \{1,2,3\}} S_i\otimes S_i \right),
\end{equation}
where $\tilde{\rho}^{(j,k)}=(U_r^{(j)}\otimes U_r^{(k)})^\dagger\rho^{(j,k)}(U_r^{(j)}\otimes U_r^{(k)})$.
Note that for spin vectors $\bm{S}^{(1)}$ and $\bm{S}^{(2)}$ acting two spin-S, we have
\begin{equation}
  \begin{aligned}
    \left[\bm{S}^{(1)}+\bm{S}^{(2)}\right]^2=&\left[S_1^{(1)}+S_1^{(2)}\right]^2+\left[S_2^{(1)}+S_2^{(2)}\right]^2+\left[S_3^{(1)}+S_3^{(2)}\right]^2\\
    =&\left[(S_1^{(1)})^2+(S_2^{(1)})^2+(S_3^{(1)})^2\right]+\left[(S_1^{(2)})^2+(S_2^{(2)})^2+(S_3^{(2)})^2\right]\\
    &+2\left[S_1^{(1)}\otimes S_1^{(2)}+S_2^{(1)}\otimes S_2^{(2)}+S_3^{(1)}\otimes S_3^{(2)}\right]\\
    =& [\bm{S}^{(1)}]^2+[\bm{S}^{(2)}]^2+2\sum_{i=1}^3 S_i^{(1)}\otimes S_i^{(2)},
  \end{aligned}
\end{equation}
thus $2\sum_{i=1}^3 S_i\otimes S_i=[\bm{S}^{(1)}+\bm{S}^{(2)}]^2-[\bm{S}^{(1)}]^2-[\bm{S}^{(2)}]^2$.
Futhermore, $[\bm{S}^{(1)}+\bm{S}^{(2)}]^2$, $[\bm{S}^{(1)}]^2$, $[\bm{S}^{(2)}]^2$ and $S_3=S_3^{(1)}+S_3^{(2)}$ can be simultaneously diagonalized using a common eigenbasis $\{\ket{S_{\text{tot}};m_{\text{tot}}}\}$ with
\begin{equation}
  \begin{aligned}
    \left[\bm{S}^{(1)}\right]^2\ket{S_{\text{tot}};m_{\text{tot}}}&=S^{(1)}(S^{(1)}+1)\ket{S_{\text{tot}};m_{\text{tot}}},\\
    \left[\bm{S}^{(2)}\right]^2\ket{S_{\text{tot}};m_{\text{tot}}}&=S^{(2)}(S^{(2)}+1)\ket{S_{\text{tot}};m_{\text{tot}}},\\
    \left[\bm{S}^{(1)}+\bm{S}^{(2)}\right]^2\ket{S_{\text{tot}};m_{\text{tot}}}&=S_{\text{tot}}(S_{\text{tot}}+1)\ket{S_{\text{tot}};m_{\text{tot}}},\\
    S_3\ket{S_{\text{tot}};m_{\text{tot}}}&=m_{\text{tot}}\ket{S_{\text{tot}};m_{\text{tot}}},
  \end{aligned}
\end{equation}
where $|S^{(1)}-S^{(2)}|\le S_{\text{tot}}\le S^{(1)}+S^{(2)}$. When the spins are all spin-S, we have $S^{(1)}=S^{(2)}=S$, $0\le S_{\text{tot}}\le 2S$ and the largest eigenvalue of $\sum_{i=1}^3 S_i\otimes S_i$ is $\frac{1}{2}\left[ 2S(2S+1)-2S(S+1) \right]=S^2$.
Thus we have $S^2 \mathbb{I}_{(2S+1)^2}-\sum_{i=1}^3 S_i\otimes S_i\geq 0$.
Note that $\tilde{\rho}^{(j,k)}$ is positive semidefinite, hence
\begin{equation}\label{eq:apprxxjk}
  \begin{aligned}
    &\tr\left( \tilde{\rho}^{(j,k)}\left[S^2 \mathbb{I}_{(2S+1)^2}-\sum_{i=1}^3 S_i\otimes S_i\right] \right)\ge 0\\
    \Rightarrow & \sum_{x\in\{\alpha,\theta,\phi\}} r^{(j,k)}_{xx}= \tr\left( \tilde{\rho}^{(j,k)}\sum_{i=1}^3 S_i\otimes S_i \right)\le S^2 \tr\left( \tilde{\rho}^{(j,k)} \mathbb{I}_{(2S+1)^2} \right)=S^2.
  \end{aligned}
\end{equation}
From these constraints, we can then obtain the ultimate lower bound as
\begin{equation}\label{eq:appgeneral}
  \begin{aligned}
    w_\alpha\delta \hat{\alpha}^{2}+w_\theta\delta \hat{\theta}^{2}+w_\phi\delta \hat{\phi}^{2}
    &\geq \frac{1}{4}\frac{\left(\sum_x \sqrt{w_x}/|c_x|\right)^2} {\left[ \sum_{k=0}^{N-1}\sum_x r^{(k)}_{xx}+\sum_{j\neq k}\sum_x r^{(j,k)}_{xx} \right]}\\
    &\geq \frac{1}{4}\frac{\left(\sum_x \sqrt{w_x}/|c_x|\right)^2} {\left[ NS(S+1)+N(N-1)S^2 \right]}\\
    &= \frac{\left(\sqrt{w_{\alpha}}+\frac{\sqrt{w_{\theta}}}{2|\sin\frac{\alpha}{2}|}+\frac{\sqrt{w_{\phi}}}{2|\sin\frac{\alpha}{2} \sin \theta|}\right)^{2}} {4NS(NS+1)}.
  \end{aligned}
\end{equation}
The bound can be saturated when
\begin{equation}
  r_{xx}^{(k)}+(N-1)r_{xx}^{(j,k)}=\frac{\frac{\sqrt{w_x}}{|c_x|}S(NS+1)}{\sum_{x\in\{\alpha,\theta,\phi\}} \frac{\sqrt{w_x}}{|c_x|}},\quad\forall x\in\{\alpha,\theta,\phi\},
\end{equation}
and $\sum_{k=0}^{N-1}r^{(k)}_\alpha=\sum_{k=0}^{N-1}r^{(k)}_\theta=\sum_{k=0}^{N-1}r^{(k)}_\phi=0$.


By employing a qutrit as the ancillary system, we can prepare the probe state as
\begin{equation}
  \ket{\Psi_0}_{SA}=P_\alpha\ket{\Phi_\alpha}_S\otimes\ket{0}_A+P_\theta\ket{\Phi_\theta}_S\otimes\ket{1}_A+P_\phi\ket{\Phi_\phi}_S\otimes\ket{2}_A,
\end{equation}
where $\ket{\Phi_x}=\frac{1}{\sqrt{2}}\left(\ket{S_x^+}^{\otimes N}+\ket{S_x^-}^{\otimes N}\right)$ for $x\in\{\alpha,\theta,\phi\}$ with $\ket{S_x^{\pm}}$ as the eigenstates corresponding to the largest and smallest eigenvalues of $S_x$, i.e. $S_x\ket{S_x^{\pm}}=\pm S\ket{S_x^{\pm}}$. The normalization condition requires that $|P_\alpha|^2+|P_\theta|^2+|P_\phi|^2=1$.


The  reduced single-spin state of this probe state is
\begin{equation}
  \rho^{(k)}=\frac{1}{2}\sum_{x\in\{\alpha,\theta,\phi\}} |P_x|^2 \left( \ket{S_x^+}\bra{S_x^+}^{(k)}+\ket{S_x^-}\bra{S_x^-}^{(k)} \right),\quad\forall 0\le k\le N-1
\end{equation}
and the  reduced two-spin state is
\begin{equation}
  \rho^{(j,k)}=\frac{1}{2}\sum_{x\in\{\alpha,\theta,\phi\}}|P_x|^2\left( \ket{S_x^+}\bra{S_x^+}^{(j)}\otimes \ket{S_x^+}\bra{S_x^+}^{(k)} +\ket{S_x^-}\bra{S_x^-}^{(j)}\otimes \ket{S_x^-}\bra{S_x^-}^{(k)} \right),\quad\forall 0\le j<k\le N-1.
\end{equation}
Using the matrix representation in Eq.(\ref{eq:appspinmatrix}), one can easily verify that
\begin{equation}
  \begin{aligned}
    \tr\left(\rho^{(k)}S_x\right)&=0,\quad\forall 0\le k\le N-1,\forall x\in\{\alpha,\theta,\phi\},\\
    \tr\left(\rho^{(j,k)}S_x\otimes S_x\right)&=|P_x|^2S^2,\quad\forall 0\le j<k\le N-1,\forall x\in\{\alpha,\theta,\phi\}.
  \end{aligned}
\end{equation}
and
\begin{equation}
  \begin{aligned}
    \tr\left(\rho^{(k)}S_\alpha^2\right)&=|P_\alpha|^2S^2+\frac{1}{2}(|P_\theta|^2+|P_\phi|^2)S,\\
    \tr\left(\rho^{(k)}S_\theta^2\right)&=|P_\theta|^2S^2+\frac{1}{2}(|P_\phi|^2+|P_\alpha|^2)S,\\
    \tr\left(\rho^{(k)}S_\phi^2\right)&=|P_\phi|^2S^2+\frac{1}{2}(|P_\alpha|^2+|P_\theta|^2)S, \quad\forall 0\le k\le N-1.
  \end{aligned}
\end{equation}

The entries of the quantum Fisher information matrix corresponding to this probe state can thus be computed as
\begin{align}
    J_{x, y \in\{\alpha, \theta, \phi\}} =&2\left\langle\Psi_{0}\left|H_{x}^{(N)} H_{y}^{(N)}+H_{y}^{(N)}H_{x}^{(N)}\right| \Psi_{0}\right\rangle \\
    &- 4\left\langle\Psi_{0}\left|H_{x}^{(N)}\right| \Psi_{0}\right\rangle\left\langle\Psi_{0}\left|H_{y}^{(N)}\right| \Psi_{0}\right\rangle, \\
    =&c_xc_y\left[2N(r^{(2)}_{xy}+r^{(2)}_{yx})+ 2N(N-1)(r_{x y}+r_{y x})-4N^2(r_{x 0}r_{y 0})\right].
\end{align}
where
\begin{equation}
  r^{(2)}_{xy}=\tr\left(\rho^{(k)}S_x^{(k)}S_y^{(k)}\right),\quad
  r_{xy}=\tr\left(\rho^{(j,k)}S_x^{(j)}\otimes S_y^{(k)}\right),\quad
  r_{x0}=\tr\left(\rho^{(k)}S_x^{(k)}\right).
\end{equation}
It can be easily checked that $r_{\alpha 0}=r_{\theta 0}=r_{\phi 0}=0$, $r^{(2)}_{\alpha\theta}+r^{(2)}_{\theta\alpha}=r^{(2)}_{\alpha\phi}+r^{(2)}_{\phi\alpha}=r^{(2)}_{\theta\phi}+r^{(2)}_{\phi\theta}=0$,  $r_{\alpha\theta}=r_{\alpha\phi}=r_{\theta\phi}=0$,
thus the quantum Fisher information matrix can be written as
\begin{equation}
    J=4N J_1+4N(N-1) J_2,
\end{equation}
where $J_1$ and $J_2$ are diagonal matrices with the diagonal entries given by
\begin{equation}
  \begin{aligned}
    \left[J_1\right]_{11}&=|P_\alpha|^2S^2+\frac{1}{2}(|P_\theta|^2+|P_\phi|^2)S,\\
    \left[J_1\right]_{22}&=4\sin^{2} \frac{\alpha}{2}\left(|P_\theta|^2S^2+\frac{1}{2}(|P_\phi|^2+|P_\alpha|^2)S\right),\\
    \left[J_1\right]_{33}&=4\sin^{2}\frac{\alpha}{2}\sin^{2} \theta \left(|P_\phi|^2S^2+\frac{1}{2}(|P_\alpha|^2+|P_\theta|^2)S\right),\\
    \left[J_2\right]_{11}&=|P_\alpha|^2S^2,\\
    \left[J_2\right]_{22}&=4\sin^{2} \frac{\alpha}{2}|P_\theta|^2S^2,\\
    \left[J_2\right]_{33}&=4\sin^{2}\frac{\alpha}{2}\sin^{2} \theta|P_\phi|^2S^2.
  \end{aligned}
\end{equation}
By substituting the QFIM in the quantum Cram\'er-Rao bound, we get
\begin{equation}
  \begin{aligned}
    \sum_{x\in\{\alpha,\theta,\phi\}} w_x\delta \hat{x}^{2}\geq & \tr\left(\operatorname{diag}\{w_\alpha,w_\theta,w_\phi\}J^{-1}\right)\\
    =& \frac{w_\alpha}{4N\left[J_1\right]_{11}+4N(N-1)\left[J_2\right]_{11}}+\frac{w_\theta}{4N\left[J_1\right]_{22}+4N(N-1)\left[J_2\right]_{22}}\\
    & +\frac{w_\phi}{4N\left[J_1\right]_{33}+4N(N-1)\left[J_2\right]_{33}}\\
    =& \frac{w_\alpha}{2NS (|P_\theta|^2+|P_\phi|^2)+4N^2S^2|P_\alpha|^2}+\frac{w_\theta/(4\sin^2\frac{\alpha}{2})}{2NS (|P_\phi|^2+|P_\alpha|^2)+4N^2S^2|P_\theta|^2}\\
    & +\frac{w_\phi/(4\sin^2\frac{\alpha}{2}\sin^2\theta)}{2NS (|P_\alpha|^2+|P_\theta|^2)+4N^2S^2|P_\phi|^2}.
  \end{aligned}
\end{equation}
It is straightforward to verify that when
\begin{equation}
  \begin{aligned}\label{eq:appgeneralcodition}
    \frac{1}{2}(|P_\theta|^2+|P_\phi|^2)+N|P_\alpha|^2S=\frac{\sqrt{w_\alpha}(NS+1)}{\sqrt{w_{\alpha}}+\frac{\sqrt{w_{\theta}}}{2|\sin\frac{\alpha}{2}|}+\frac{\sqrt{w_{\phi}}}{2|\sin\frac{\alpha}{2} \sin \theta|}},\\
    \frac{1}{2}(|P_\phi|^2+|P_\alpha|^2)+N|P_\theta|^2S=\frac{\frac{\sqrt{w_{\theta}}}{2|\sin\frac{\alpha}{2}|}(NS+1)}{\sqrt{w_{\alpha}}+\frac{\sqrt{w_{\theta}}}{2|\sin\frac{\alpha}{2}|}+\frac{\sqrt{w_{\phi}}}{2|\sin\frac{\alpha}{2} \sin \theta|}},\\
    \frac{1}{2}(|P_\alpha|^2+|P_\theta|^2)+N|P_\phi|^2S=\frac{\frac{\sqrt{w_{\phi}}}{2|\sin\frac{\alpha}{2} \sin \theta|}(NS+1)}{\sqrt{w_{\alpha}}+\frac{\sqrt{w_{\theta}}}{2|\sin\frac{\alpha}{2}|}+\frac{\sqrt{w_{\phi}}}{2|\sin\frac{\alpha}{2} \sin \theta|}},
  \end{aligned}
\end{equation}
it achieves the ultimate lower bound given in Eq.(\ref{eq:appgeneral}), which is
\begin{equation}
  \begin{aligned}
    \sum_{x\in\{\alpha,\theta,\phi\}} w_x\delta \hat{x}^{2}\geq \frac{\left(\sqrt{w_{\alpha}}+\frac{\sqrt{w_{\theta}}}{2|\sin\frac{\alpha}{2}|}+\frac{\sqrt{w_{\phi}}}{2|\sin\frac{\alpha}{2} \sin \theta|}\right)^{2}} {4NS(NS+1)}.
  \end{aligned}
\end{equation}
The optimal coefficient can be obtained by solving Eq.(\ref{eq:appgeneralcodition}) explicitly as
\begin{equation}
  \begin{aligned}
    |P_\alpha|^2&=\frac{(2NS+1)\sqrt{w_{\alpha}}-\frac{\sqrt{w_{\theta}}}{2|\sin\frac{\alpha}{2}|}-\frac{\sqrt{w_{\phi}}}{2|\sin\frac{\alpha}{2} \sin \theta|}}{(2NS-1)(\sqrt{w_{\alpha}}+\frac{\sqrt{w_{\theta}}}{2|\sin\frac{\alpha}{2}|}+\frac{\sqrt{w_{\phi}}}{2|\sin\frac{\alpha}{2} \sin \theta|})},\\
    |P_\theta|^2&=\frac{(2NS+1)\frac{\sqrt{w_{\theta}}}{2|\sin\frac{\alpha}{2}|}-\frac{\sqrt{w_{\phi}}}{2|\sin\frac{\alpha}{2} \sin \theta|}-\sqrt{w_{\alpha}}}{(2NS-1)(\sqrt{w_{\alpha}}+\frac{\sqrt{w_{\theta}}}{2|\sin\frac{\alpha}{2}|}+\frac{\sqrt{w_{\phi}}}{2|\sin\frac{\alpha}{2} \sin \theta|})},\\
    |P_\phi|^2&=\frac{(2NS+1)\frac{\sqrt{w_{\phi}}}{2|\sin\frac{\alpha}{2} \sin \theta|}-\sqrt{w_{\alpha}}-\frac{\sqrt{w_{\theta}}}{2|\sin\frac{\alpha}{2}|}}{(2NS-1)(\sqrt{w_{\alpha}}+\frac{\sqrt{w_{\theta}}}{2|\sin\frac{\alpha}{2}|}+\frac{\sqrt{w_{\phi}}}{2|\sin\frac{\alpha}{2} \sin \theta|})}.
  \end{aligned}
\end{equation}
If the right-hand-sides of the above equations are all non-negative, then the coefficients of the optimal probe state can be obtained by taking the square root. This is always the case for sufficiently large $N$ or $S$. The ultimate lower bound in Eq.(\ref{eq:appgeneral}) can thus always be saturated for sufficiently large $N$ or $S$.

It is also straightforward to check that the weak commutativity conditions are satisfied, i.e.
\begin{equation}
  \begin{aligned}
    &\mathrm{Im} \left<\partial_x\Phi(\alpha,\theta,\phi)|\partial_y\Phi(\alpha,\theta,\phi)\right>\\
    =&\mathrm{Im} \bra{\Psi_{SA}}(H_x^{(N)})(H_y^{(N)})\ket{\Psi_{SA}}\\
    =& \mathrm{Im} \left[ N\bra{\Psi_{SA}}H_x^{[1]}H_y^{[1]}\ket{\Psi_{SA}}+N(N-1)c_x c_y r_{xy} \right]\\
    =&N c_x c_y \epsilon_{xyz}r_{z0}\\
    =&0,
  \end{aligned}
\end{equation}
where $x,y,z$ represent three different parameters in $\{\theta,\phi,\alpha\}$ and $\epsilon_{x y z}$ is the Levi-Civita symbol if we label $(\theta, \phi, \alpha)$ as $(1,2,3)$ respectively, and
in the last two equalities we have used the facts that for the chosen probe state
\begin{equation}
  r_{z 0}=\tr\left(\rho^{(k)}S_z\right)=0,\forall z\in\{\alpha,\theta,\phi\},
\end{equation}
\begin{equation}
  r_{x y}=\tr\left(\rho^{(j,k)}S_x\otimes S_y\right)=0,\forall x\neq y\in\{\alpha,\theta,\phi\}.
\end{equation}
\section{Weights for $\sum_{i=1}^{3}\delta\hat{B}_i^2$}
When the figure of merit is taken as $\sum_{i=1}^{3}\delta\hat{B}_i^2$, the weights in the representation of $\{\alpha, \theta,\phi\}$ can be obtained from the error propagation formula. Since $B_1=\frac{\alpha}{t}\sin\theta\cos\phi$, $B_2=\frac{\alpha}{t}\sin\theta\sin\phi$, $B_3=\frac{\alpha}{t}\cos\theta$, we have
\begin{eqnarray}
\nonumber
\aligned
\delta \hat{B}_1&=\sin\theta\cos\phi\frac{\delta \hat{\alpha}}{t}+\frac{\alpha}{t}\cos\theta\cos\phi\delta\hat{\theta}-\frac{\alpha}{t}\sin\theta\sin\phi\delta\hat{\phi},\\
\delta \hat{B}_2&=\sin\theta\sin\phi\frac{\delta \hat{\alpha}}{t}+\frac{\alpha}{t}\cos\theta\sin\phi\delta\hat{\theta}+\frac{\alpha}{t}\sin\theta\cos\phi\delta\hat{\phi},\\
\delta \hat{B}_3&=\cos\theta\frac{\delta \hat{\alpha}}{t}-\frac{\alpha}{t}\sin\theta\delta\hat{\theta}.\\
\endaligned
\end{eqnarray}
It is then straightforward to get
\begin{eqnarray}
\delta \hat{B}_1^2+\delta \hat{B}_2^2+\delta \hat{B}_3^2=\frac{\delta\hat{\alpha}^2}{t^2}+\frac{\alpha^2}{t^2}\delta\hat{\theta}^2+\frac{\alpha^2\sin^2\theta}{t^2}\delta\hat{\phi}^2.
\end{eqnarray}
\end{widetext}
\end{document}